\documentclass[11pt]{article}
\pdfoutput=1 
\usepackage{amsmath,amssymb}
\usepackage{jheppub}
\usepackage{bm,mathtools,slashed}
\usepackage{physics}
\usepackage{mathrsfs}
\usepackage{babel}
\newcommand{\rmi}{\mathrm{i}}
\newcommand{\rme}{\mathrm{e}}
\newcommand{\rmd}{\mathrm{d}}
\newcommand{\calL}{\mathcal{L}}
\newcommand{\calN}{\mathcal{N}}
\newcommand{\intperp}{\int\rmd^2p_\perp}

\newcommand{\nus}{\nu_\mathrm{S}}
\newcommand{\nud}{\nu_\mathrm{D}}

\newcommand{\bp}{{\bm p}}
\newcommand{\bq}{{\bm q}}
\newcommand{\bx}{{\bm x}}
\newcommand{\by}{{\bm y}}
\newcommand{\bz}{{\bm z}}

\newcommand{\Da}{D^\mathrm{a}}
\newcommand{\Sa}{S^\mathrm{a}}

\newcommand{\hatSS}{\hat{\mathscr{U}}_\mathrm{S}}
\newcommand{\hatSD}{\hat{\mathscr{U}}_\mathrm{D}}
\newcommand{\hataout}{\hat{a}_{\mathrm{out},\bp}}
\newcommand{\hatbout}{\hat{b}_{\mathrm{out},-\bp}}

\newcommand{\hataass}{\hat{a}_{\mathrm{as},\bp}^{(s)}}
\newcommand{\hatbass}{\hat{b}_{\mathrm{as},-\bp}^{(s)}}
\newcommand{\hataoutsdag}{\hat{a}_{\mathrm{out},\bp}^{(s)\dag}}
\newcommand{\hatboutsdag}{\hat{b}_{\mathrm{out},-\bp}^{(s)\dag}}

\newcommand{\inket}{|0_\mathrm{in}\rangle}
\newcommand{\outket}{|0_\mathrm{out}\rangle}
\newcommand{\inbra}{\langle0_\mathrm{in}|}
\newcommand{\outbra}{\langle0_\mathrm{out}|}

\newcommand{\alp}{\alpha_\bp}
\newcommand{\betap}{\beta_\bp}
\newcommand{\sigmap}{\sigma_\bp}

\newcommand{\Upsas}[2]{U_{{\bm #1},#2}^\text{as}}
\newcommand{\Vpsas}[2]{V_{{\bm #1},#2}^\text{as}}

\newcommand{\Upas}[1]{U_{{\bm #1}}^\text{as}}
\newcommand{\Vpas}[1]{V_{{\bm #1}}^\text{as}}
\newcommand{\Upsin}[2]{U_{{\bm #1}#2}^\text{in}}
\newcommand{\Vpsin}[2]{V_{{\bm #1}#2}^\text{in}}
\newcommand{\bUpsin}[2]{\bar{U}_{{\bm #1}#2}^\text{in}}

\newcommand{\Upin}[1]{U_{{\bm #1}}^\text{in}}
\newcommand{\Vpin}[1]{V_{{\bm #1}}^\text{in}}
\newcommand{\Upsout}[2]{U_{{\bm #1}#2}^\text{out}}
\newcommand{\Vpsout}[2]{V_{{\bm #1}#2}^\text{out}}

\newcommand{\bVpsout}[2]{\bar{V}_{{\bm #1}#2}^\text{out}}
\newcommand{\Upout}[1]{U_{{\bm #1}}^\text{out}}
\newcommand{\Vpout}[1]{V_{{\bm #1}}^\text{out}}
\title{In-in formalism with resummmation in a constant electric field: propagators including nontrivial boundary wavefunctions}

\preprint{}

\author{Kenji Fukushima and Shuhei Minato}

\affiliation{Department of Physics, The University of Tokyo, 7-3-1 Hongo, Bunkyo-ku, Tokyo 113-0033, Japan}

\emailAdd{fuku@nt.phys.s.u-tokyo.ac.jp}
\emailAdd{minato@nt.phys.s.u-tokyo.ac.jp}

\abstract{
We present the derivation of an alternative representation of the real-time in-in formalism under a spatially homogeneous and time independent electric field.
Because the system exhibits instability associated with pair production of particles and antiparticles, the perturbation theory should be reorganized depending on the choice of the reference vacuum.
We recast the boundary wavefunctions into the quadratic self-energy-like terms in the functional integration formalism. 
The resulting generating functional in the modified in-in formalism leads to the propagators that resum infinite diagrams necessary to capture the vacuum-instability effects.
The proper-time representations of the propagators reproduce the known expressions from the canonical operator formalism, but our derivation based on the generating functional along the closed-time path clarifies the origin of the additional proper-time contour and provides a better physical understanding.
Finally, as a concrete example of the application, we compute the in-in expectation value of the vector current in a constant electric field, and find that the simple one-loop calculation captures the pair production effect. 
}
\begin{document}
\maketitle

%%%%%%%%%%%%%%%%%%%%%%%%%%%%%%%%%%%%%%%%%
\section{Introduction}
%%%%%%%%%%%%%%%%%%%%%%%%%%%%%%%%%%%%%%%%%

Strong external electric fields induce intrinsically nonperturbative real-time phenomena in
quantum field theory, most prominently Schwinger pair production~\cite{Schwinger:1951nm} and the
associated vacuum instability~\cite{Hebenstreit:2009km,Kluger:1992md,Blaschke:2008wf,Fedotov:2010ja,Elkina:2010up,Dumlu:2010ua,Tanji:2008ku,Schutzhold:2008pz,Taya:2020bcd}.
Such strong-field environments are not merely of conceptual interest: ultra-strong electromagnetic
fields are created in heavy-ion collisions~\cite{Kharzeev:2007jp,Voronyuk:2011jd,Huang:2015oca,ATLAS:2017fur,ATLAS:2019azn,CMS:2018erd,STAR:2019wlg},
are targeted in high-intensity laser facilities~\cite{DiPiazza:2011tq,Fedotov:2022ely}, and may also appear in compact astrophysical systems~\cite{Enoto:2019vcg}. 
Also, recently, the system with acceleration, which causes the vacuum instability analogous to the electric field effects, is attracting interest~\cite{Ohsaku:2004rv,Benic:2015qha,Becattini:2020qol,Ambrus:2023smm,Chernodub:2025ovo}. 
A quantitative description of these settings
requires a framework that can follow nonequilibrium evolution in real time while properly encoding
vacuum instability.

The in-in (or the closed-time-path or the Schwinger-Keldysh) formalism is the standard real-time framework in
quantum field theory~\cite{Schwinger:1960qe,Keldysh:1964ud,Berges:2004yj}. In practice, one often builds perturbation theory from the
in-out propagators with the usual Feynman boundary conditions implemented by the $\rmi\epsilon$
prescription~\cite{Weinberg:2005vy,Calzetta:2008iqa}. This choice is adequate for stable vacua, but in a background
electric field it does not by itself capture the physics of vacuum decay. Observables directly tied
to vacuum instability (e.g.\ pair production and the induced current) then require a consistent
resummation of infinitely many diagrams~\cite{Fedotov:2022ely,Copinger:2025ovz}. The issue is not only technical: when one
tests operator relations such as Ward identities or energy-balance relations in real time, both
sides must be resummed on the same footing~\cite{Copinger:2018ftr}, which makes a straightforward loop
expansion cumbersome and can obscure the physical interpretation.

A complementary viewpoint exists in the canonical operator formalism. There one can work with
propagators that already incorporate the vacuum-instability effects of external fields and that can
be expressed in proper-time representation~\cite{Nikishov:1969tt,Fradkin:1991zq}. With these propagators,
real-time quantities such as the induced current can be computed at a fixed loop order without any
additional diagrammatic resummation~\cite{Copinger:2018ftr,Gavrilov:2007hq,Gavrilov:2012jk}. Despite their usefulness,
however, it has remained conceptually unclear how to embed such resummed propagators \emph{naturally}
into the functional-integral formulation of the in-in formalism, i.e., how the required resummation
should arise from the boundary conditions of the closed-time path.

In this work we provide such an embedding. The key observation is that, in a constant electric
field, the definition of particles and vacuum is not unique: the in- and out-vacua are inequivalent
and related by a Bogoliubov transformation. This fact implies that the ``in-in boundary condition''
admits distinct but {equally valid} representations. We formulate the in-in generating functional so as
to keep the boundary wavefunctions explicitly, and then reorganize the perturbative expansion by
choosing the \emph{out}-state as the reference vacuum. In the functional integral, the boundary
wavefunctions can be recast into quadratic, boundary-localized kernels that enter the action as
self-energy--like terms. Importantly, this reorganization leaves the interaction vertices intact
and modifies only the quadratic part, so that the resulting real-time propagators are obtained from the Dyson series dictated by the boundary condition.

This formulation has some advantages. First, it provides a practical framework in which
vacuum-instability effects are incorporated at the propagator level, so in-in expectation values
can be computed without tracking an explicit infinite resummation. Second, it clarifies the
proper-time structure of the resummed propagators: while proper-time expressions in an electric
field are known in canonical treatments, the appearance of additional proper-time contours may look mystical. By deriving the propagators directly from the closed-time-path generating functional,
we identify the origin of these additional proper-time contours and give them a transparent
interpretation in terms of the nontrivial boundary wavefunctions associated with the unstable vacuum.

As a concrete demonstration, we compute the in-in expectation value of the vector current in a
constant electric field at one loop using the modified propagators. The resummation is then
automatically included, and the resulting expression exhibits the physically expected time-growing
behavior associated with Schwinger pair production. Beyond this illustrative application, the
functional framework is well suited for nonperturbative mean-field analyses~\cite{Andersen:2014xxa,Miransky:2015ava,Cao:2021rwx,Hattori:2023egw},
and it offers a useful perspective for attempts to understand real-time electric-field physics with
lattice studies employing real or imaginary electric fields~\cite{Yamamoto:2012bd,Yamamoto:2021oys,Yang:2023zzx,Endrodi:2022wym,Endrodi:2023wwf,Endrodi:2024cqn},
where questions of analytic continuation and its limitations are central. 

This paper is organized as follows. In Sec.~\ref{sec:unstable} we review the
standard real-time propagators in a constant electric field and pinpoint why the na\"{i}ve in-in setup
built from in-out propagators fails to capture vacuum-instability observables without resummation.
In Sec.~\ref{sec:recast} we discuss the role of inequivalent vacua and reformulate
the real-time expectation values with the effects of nontrivial boundary wavefunctions. In Sec.~\ref{sec:generating}
we construct the corresponding generating functionals on the closed-time path. In
Sec.~\ref{sec:propagator} we derive the resulting resummed real-time propagators and their
proper-time representations, elucidating the origin of the additional proper-time contours. In
Sec.~\ref{sec:application} we present the one-loop computation of the vector current as an
application. Section~\ref{sec:conclusion} contains our conclusions and outlook.

For definiteness, we consider a spatially homogeneous and temporally constant electric field and
ignore magnetic fields. We assume that interactions are adiabatically switched off in the
asymptotic past and future. Throughout this paper we use natural units, $\hbar=1=c$, and adopt the
metric convention $g_{\mu\nu}=\mathrm{diag}(+,-,-,-)$. In addition, we use the following shorthand notations: $\int_{\bp}:=\int\frac{\rmd^3p}{(2\pi)^3},\quad \int_{\bx}:=\int \rmd^3x$.

%%%%%%%%%%%%%%%%%%%%
\section{Real-time propagators between unstable vacua}
\label{sec:unstable}
%%%%%%%%%%%%%%%%%%%%

It is widely known that Schwinger derived a proper-time expression of propagators in the presence of general electromagnetic background fields~\cite{Schwinger:1951nm}.  For applications to a constant magnetic field, even today, Schwinger's propagator is preferably adopted as the most convenient tool in practice (see Ref.~\cite{Miransky:2015ava} for a review).  In the case of electric field, in contrast, the calculation and interpretation encounter some subtleties as discussed below.  The Dirac propagator under a constant electric field takes the form of
\begin{align}
  S_0(x,y) =
  \langle x| \frac{\rmi (\rmi \slashed{D} + m)}{-\slashed{D}^2 - m^2 + \rmi\epsilon} |y\rangle
  = \rmi (\rmi \slashed{D}_x + m) \int_{\Gamma_\mathrm{c}} \rmd s\; g(s;x,y)\,,
    \label{eq:S}
\end{align}
where
\begin{equation}
  g(s;x,y) :=
  \langle x| \rme^{\rmi (-\slashed{D}^2 - m^2 + \rmi\epsilon)} |y\rangle
  = \rme^{\gamma^0\gamma^3 eEs}\,f(s;x,y)
\end{equation}
and $f(s;x,y)$ is a factorized spinless part that is given explicitly by
\begin{align}
  \begin{split}
    f(s;x,y) &:= \frac{eE}{(4\pi)^2 \, s\, \sinh(eEs)} \times \\
    &\quad \times \exp\biggl[ \rmi\frac{eE}{2} (x_0+y_0)z_3
    - \rmi\frac{eE}{4} (z_0^2-z_3^2)\coth(eEs)
    + \rmi\,\frac{{\bm z}_\perp^2}{4s} - \rmi\, (m^2-\rmi\epsilon)s  \biggr]\,.
  \end{split}
  \label{eq:f}
\end{align}
Here, the relative coordinate is $z:=x-y$ and a shorthand notation, $\bz_\perp^2 := z_1^2 + z_2^2$, is introduced.  Throughout this work, the electric field is imposed along the $x^3$ axis with $E>0$.  Thus, the vector potential is chosen to be
\begin{align}
    A^\mu(x) = (0,0,0,-E x^0)\,,
    \label{eq:vector-potential}
\end{align}
where $E>0$. We note that the first term in the exponential appears from the Schwinger phase, i.e.,
\begin{align}
    \exp\biggl(\rmi e\int_x^y \rmd \ell \cdot A \biggr)
    = \exp\biggl[\rmi  \frac{eE}{2}(x_0 + y_0) z_3\biggr]
\end{align}
under the choice of the straight line path from $x$ to $y$.

%--- figure ---%
\begin{figure}
    \centering
    \includegraphics[width=0.3\textwidth]{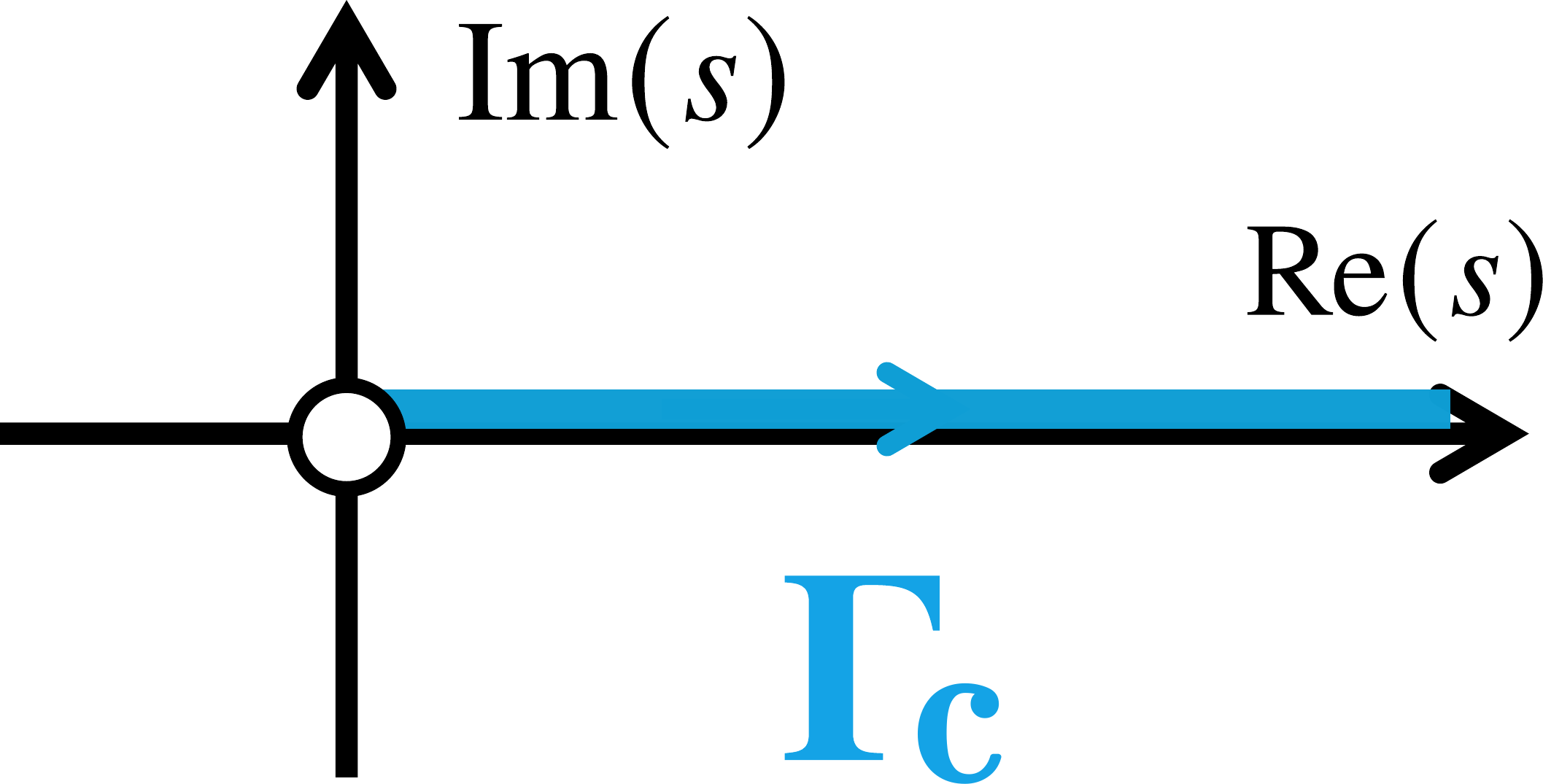}
    \caption{Proper-time path in the standard Schwinger propagator.}
    \label{fig:Gammac}
\end{figure}
%--- figure ---%

Here, it is important to make a remark that the proper-time contour, $\Gamma_\mathrm{c}$, extends from $s=0$ to $s=+\infty$ along the real axis as shown in Fig.~\ref{fig:Gammac}.  If this propagator in Eq.~\eqref{eq:S} is na\"{i}vely applied to perturbative calculations, the leading-order results may become different from what is expected from physical intuitions.  For example, consider the current along the electric field, $\langle \hat{j}^3(x^0)\rangle = \langle\bar{\psi}\gamma^3\psi\rangle$. One expects it to be nonzero due to pair production and acceleration under the electric field.  However, if one employs $S_0(x,y)$ in Eq.~\eqref{eq:S} to calculate $\langle\hat{j}^3(x^0)\rangle$ at one-loop level, one would find $\langle \hat{j}^3(x^0)\rangle = 0$~\cite{Fradkin:1991zq,Gavrilov:1998hw,Copinger:2018ftr}.  Although it is not a preferable answer in physics, there is no apparent flaw in the mathematical formulation.

\begin{figure}
    \centering
    \includegraphics[width=0.75\textwidth]{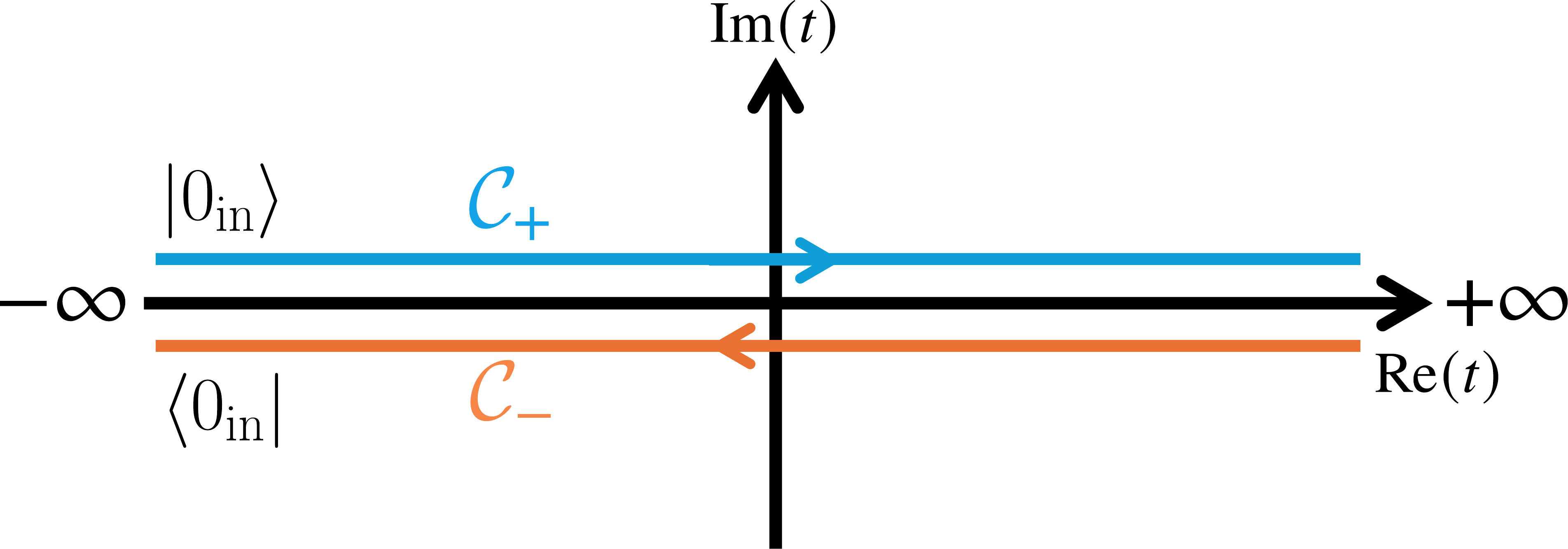}
    \caption{Time-integral contour in the standard in-in formalism.  The in-vacuum at the infinite past evolves to the infinite future, and then returns to the infinite past.}
    \label{fig:inin}
\end{figure}

It is a nontrivial question where to identify any potential drawback in the above calculation machinery.  Let us shed a light by considering this problem in the closed-time path or the in-in formalism.
To illustrate the setup of the standard in-in formalism, we take the time-integral contour as depicted in Fig.~\ref{fig:inin}.
In the asymptotically infinite past and future, the interaction is adiabatically turned off.  Accordingly, one can define the vacua and the particles in those asymptotic regions.  Following the ordinary convention, we call the vacuum in the infinite past the ``in-vacuum'' denoted by $\inket$, and the vacuum in the infinite future the ``out-vacuum'' denoted by $\outket$.

%--- figure ---%
%--- figure ---%

In the in-in formalism~\cite{Weinberg:2005vy,Calzetta:2008iqa}, one computes the expectation value of some time-dependent operator in the Heisenberg picture using the $\mathcal{C}_+$ path (upper path) and the $\mathcal{C}_-$ path (lower path) along the closed-time path in Fig.~\ref{fig:inin}.  Because the action takes different forms on the upper and the lower branches of the path, the in-in propagator, $S(x,y)$, {has} four independent components, namely, $S^{11}(x,y)$, $S^{22}(x,y)$, $S^{12}(x,y)$, and $S^{21}(x,y)$.
% We note that the bar is attached to distinguish the na\"{i}ve propagators from the resummed ones.
To determine the propagators uniquely in general, one must specify the boundary conditions.  
In the path-integral language, the boundary condition and the temporal evolution correspond to the boundary wavefunction and the Feynman kernel, respectively.  
For the insertion of the in-vacuum wavefunction, as discussed in Refs.~\cite{Weinberg:1995mt,Weinberg:2005vy}, one does not have to write down the wavefunction explicitly in the $\rmi\epsilon$ prescription.  
In the na\"{i}ve implementation of the in-in formalism, the $2\times 2$ propagators are written as~\cite{Berges:2004yj,Weinberg:2005vy,Calzetta:2008iqa}:
\begin{subequations}
\begin{align}
    & S^{11}(x,y) = S_0(x,y)\,,\\
    & S^{22}(x,y) = \gamma^0 \big[S^{11}_0(y,x)\bigr]^\dag\gamma^0\,,\\
    & S^{21}(x,y) = S^{>}_0(x,y)\,,\\
    & S^{12}(x,y) = -S^{<}_0(x,y)\,.
\end{align}
\label{eq:naive-propagator}
\end{subequations}
It is crucial to note that $S^{11}(x,y)$ is the standard in-out propagator in the $\rmi\epsilon$ prescription, that is nothing but $S_0(x,y)$.  
The off-diagonal components are found from the decomposition of
\begin{align}
    S_0(x,y) = \theta(z_0)\,S^{>}_0(x,y) - \theta(-z_0)\,S^{<}_0(x,y)\,.
\end{align}
As already mentioned, the one-loop calculation in terms of the above propagators~\eqref{eq:naive-propagator} leads to $\langle \hat{j}^3(x^0)\rangle = 0$ even in the in-in formalism.  In other words, one must take some proper resummation of diagrams to reach a physically natural conclusion of $\langle \hat{j}^3(x^0)\rangle \neq 0$.  The resummation program is in principle feasible, but it would need delicate treatments.  
The following example could better explain this point.  Suppose that one is interested in the energy balance relation, i.e.,
\begin{align}
    \partial_\mu \hat{T}^{\mu0} = F^{0\mu}\hat{j}_\mu \,.
\end{align}
Then it would be a nontrivial question how to maintain consistency in taking the expectation values of the left and right sides.  
In a uniform system, $\partial_i \langle\hat{T}^{i\nu}\rangle=0$ holds, and the energy conservation law is satisfied with the Joule heat contribution, resulting in a sensible form:
\begin{align}
    \partial_0 \langle\hat{T}^{00}\rangle = E\langle\hat{j}^3\rangle\,.
\end{align}
To incorporate the effects of pair production into these expectation values of $\langle\hat{T}^{00}\rangle$ and $\langle\hat{j}^3\rangle$, one must take the resummation of diagrams on the left and right sides on an equal footing.  
It is important to evaluate these two physical quantities in a consistent way corresponding to the common wavefunction and time evolution.  
This is a highly nontrivial task, and therefore, it would be desirable to develop an alternative framework that is more suited for the pair production problem.

Such a well-suited framework has been constructed in the operator formalism~\cite{Nikishov:1969tt,Fradkin:1991zq}.
Using the modified propagators, one can compute the current expectation value at the simplest one-loop order to find
\begin{align}
    \langle  \hat{j}^3(x^0) \rangle = \frac{(eE)^2 x^0}{2\pi^3}\, \rme^{-\pi m^2 / (eE)} \,,
    \label{eq:j3-calc}
\end{align}
which finds the well-known formula under the name of the Schwinger mechanism.
We will give a detailed derivation of this formula later in Sec.~\ref{sec:application}.

As described above, in the standard in-in formalism using the propagators in Eq.~\eqref{eq:naive-propagator}, it is technically difficult to derive Eq.~\eqref{eq:j3-calc}, because this is essentially a nonperturbative tunneling phenomenon, and the resummation of infinite number of diagrams would be required to discuss such phenomena related to the vacuum instability.  Since the in-in formalism takes the most natural form in the functional integration representation, it would be an interesting challenge to implement a more convenient method to derive the results such as Eq.~\eqref{eq:j3-calc} using not the operator but the functional integration formalism.

%%%%%%%%%%%%%%%%%%%%
\section{Recasting the vacuum with boundary wavefunction}
\label{sec:recast}
%%%%%%%%%%%%%%%%%%%%

Before presenting detailed expressions, we shall clarify the meaning of two different schemes in the standard in-in and resummed formalisms within the same setup.  For the moment, we continue our discussions for the Dirac field under a spatially homogeneous electric field with the vector potential in Eq.~\eqref{eq:vector-potential}.  Later, we will derive various formulas for the complex scalar field as well as the Dirac field.

In the standard in-in formalism, it is the wavefunction at the boundary in the infinite past, $\inket$, that determines the propagators.  It is the crucial observation that, under the electric field, the in-vacuum and the out-vacuum are not equivalent in association with the vacuum instability.  In fact, these two vacua are related through the Bogoliubov transformation as follows~\cite{Fradkin:1991zq,Kim:2008yt}:
\begin{align}
    \inket = \calN \hatSD^\dag \, \outket\,,
    \qquad
    \outket = \calN' \hatSD \, \inket\,,
    \label{eq:vac-transform}
\end{align}
where $\calN$ and $\calN'$ are normalization constants.  The normalization condition leads to $|\calN|^2 = \outbra \hatSD \hatSD^\dag \outket^{-1}$.  We emphasize that, even if the vacua are properly normalized, this normalization factor is nontrivial because of the nonunitary nature of the Bogoliubov transformation which reflects the vacuum instability.  The Bogoliubov transformation reads:
\begin{equation}
    \hatSD^\dag := \exp\biggl[ \sum_s \int_\bp\, \sigma_\bp^\ast\; \hataoutsdag\, \hatboutsdag \biggr] \,,
    \qquad
    \sigma_\bp := \frac{\betap}{\alp}
    \label{eq:hatSD_exp}
\end{equation}
with $\alp$ and $\betap$ the Bogoliubov coefficients fixed from the asymptotic behavior of the solutions of the equation of motion.  We summarize the explicit expressions of $\alp$ and $\betap$ in Appendix~\ref{app:mode-function}.
Because we can represent $\inket$ differently by means of the transformation~\eqref{eq:vac-transform}, we can reorganize the perturbation theory with different forms of the propagators, which is the most important point.

Thus, using Eq.~\eqref{eq:vac-transform}, one can rewrite the in-in expectation value of an operator $\hat{O}$ as
\begin{align}
    \inbra \hat{O}(x^0) \inket = |\calN|^2  \outbra \hatSD\, \hat{O}(x^0)\, \hatSD^\dag \outket \,.
    \label{eq:out-out-exp-wovb}
\end{align}
If the right-hand side is diagrammatically expanded, $\hatSD$ and $\hatSD^\dag$ generate vacuum bubbles but they are subtracted by $|\calN|^2$. 
We stress that, in evaluating the operator expectation values in the na\"{i}ve loop expansion, the equality~\eqref{eq:out-out-exp-wovb} may not hold at each expansion order.
Importantly, the leading-order result in the loop expansion on the out-states is regarded as the one after summing infinite diagrams in the loop expansion on the in-states.

\begin{figure}
    \centering
    \includegraphics[width=0.75\linewidth]{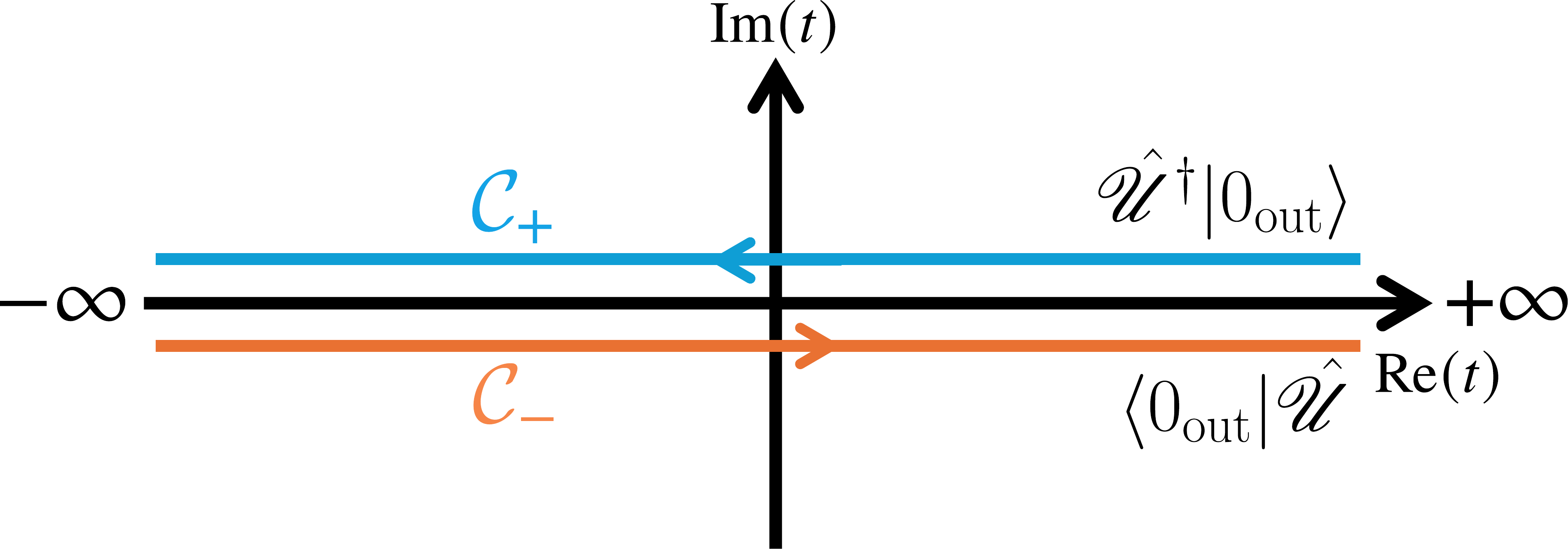}
    \caption{Modified time-integral contour in the resummed in-in formalism for a constant electric field.  The boundary state is transformed according to Eq.~\eqref{eq:vac-transform}, and the time contour starts from the infinite future to the infinite past.}
    \label{fig:outout}
\end{figure}

With these remarks, it would be conceivable that the two sides of Eq.~\eqref{eq:out-out-exp-wovb} may lead to different representations of the same operator expectation value.  The boundary conditions for these quantities in the functional integration representation should differ significantly; the left-hand side in the closed-time path from $\inket$ to $\inbra$ corresponds to the boundary condition leading to the propagators in Eq.~\eqref{eq:naive-propagator} as shown in Fig.~\ref{fig:inin}, whereas the closed-time path in the right-hand side extends from $\outket$ to $\outbra$ with the effects of $\hatSD$ and $\hatSD^\dag$; see Fig.~\ref{fig:outout}.  Thus, the latter yields the propagators different from Eq.~\eqref{eq:naive-propagator}.
In the following, we will see that $\hatSD$ and $\hatSD^\dag$ give rise to an effective self-energy term, and the modified propagators are obtained as a result of resummation in an explicit way (see Appendix \ref{app:derivation} for the explicit calculation).

%--- figure ---%
%--- figure ---%

%%%%%%%%%%%%%%%%%%%%
\section{Generating functionals}
\label{sec:generating}
%%%%%%%%%%%%%%%%%%%%

We shall exploit an extended formalism of the generating functional and derive the integral forms of the boundary wavefunctions.  We will consider scalar quantum electrodynamics (sQED) with a complex scalar field defined by the Lagrangian density:
\begin{align}
    \calL = |D_\mu \hat{\phi}|^2 - m^2|\hat{\phi}|^2 - V_{\mathrm{S}}(\hat{\phi}, \hat{\phi}^\dag) \,,
    \label{eq:Lag_scalar}
\end{align}
where the covariant derivative is $D_\mu = \partial_\mu + \rmi eA_\mu$ with the external field $A_\mu$ in Eq.~\eqref{eq:vector-potential}.
%In our convention, we use $\hat{\phi}$ for the operator of the complex scalar field, while $\phi$ denotes the integration variable.
We also discuss fermionic quantum electrodynamics (called QED simply in this paper) with a Dirac fermion field defined by the Lagrangian density:
\begin{align}
    \calL = \bar{\psi}(\rmi\slashed{D} -m)\psi - V_{\mathrm{D}}(\psi,\bar{\psi}) \,.
    \label{eq:Lag_dirac}
\end{align}
For notational brevity, we omit the hat for the Dirac field (and for the scalar field as long as no confusion arises).  In this paper, we will not look into higher-order perturbative contributions.  Thus, the explicit forms of the scalar interaction $V_{\mathrm{S}}(\phi,\phi^\dag)$ and the fermionic interaction $V_{\mathrm{D}}(\psi,\bar{\psi})$ are irrelevant in our present discussions.  It is not difficult to apply the present formalism to the linear sigma model or the Nambu--Jona-Lasinio model if the interaction effects play a significant role.

%%%%%%%%%%
\subsection{Formal expressions of the generating functionals}

In the standard functional-integration formulation of the in-in generating functional, the boundary wavefunction is a trivial Gaussian under the assumption that only the ground state should be dominant in the infinite past. 
By contrast, in the presence of time-dependent background fields, the in-state and the out-state are inequivalent, and the definition of the generating functional which we derive below retains the discrepancy between these two states through the insertion of operators $\hat{\mathscr{U}}_{\rm S/D}$ and $\hat{\mathscr{U}}^\dagger_{\rm S/D}$ at the boundary. 
Due to these operators, the boundary wavefunction becomes nontrivial (non-Gaussian), which induces the nonlocal self-energy-like terms in the generating functional and thereby modifies the propagators.

We are interested in the in-in propagator rewritten in terms of the out-state, so that the convenient choice of the time contour should be the path starting from $x^0=+\infty$, going through $x^0=-\infty$ on $\mathcal{C}_+$, and returning back to $x^0=+\infty$ on $\mathcal{C}_-$, as illustrated in Fig.~\ref{fig:outout}.
We could have renamed $\mathcal{C}_\pm \to \mathcal{C}_\mp$, but here, we intentionally keep our convention of $\mathcal{C}_\pm$ to make a clear contrast between Figs.~\ref{fig:inin} and \ref{fig:outout}.
Apart from the time contour, we will follow the basic procedures to write down the generating functional~\cite{Berges:2004yj,Weinberg:2005vy,Calzetta:2008iqa}.

%%%%%
\paragraph{sQED:}

The generating functional for the complex scalar field $\phi$ and its complex conjugate $\phi^\ast$ in sQED is
\begin{align}
    Z[J,J^*] &= \int[\rmd\phi_+][\rmd\phi_+^\ast] \; \Psi_{\mathrm{S}}(\phi_+(+\infty))\; \exp\Big[ -\rmi S(\phi_+,\phi_+^\ast) - \rmi\int \rmd^4x\,(J_+\phi_+ + J_+^\ast\phi_+^\ast)\Big] \times \notag\\
    &\quad\times \int [\rmd\phi_-][\rmd\phi^*_-]\; \Psi_{\mathrm{S}}^\ast(\phi_-(+\infty))\; \exp\Big[ +\rmi S(\phi_-,\phi_-^\ast) + \rmi\int \rmd^4x\,(J_-\phi_- + J_-^\ast\phi_-^\ast)\Big] \times \notag\\
    &\quad\times \delta\big[\phi_+(-\infty) - \phi_-(-\infty)\big] \, \delta\big[\phi^\ast_+(-\infty) - \phi_-^\ast(-\infty)\big]\,,
    \label{eq:PathInt_scalar}
\end{align}
where $\phi_+(+\infty)$ is a shorthand notation for $\phi_+(x^0=+\infty,\bx)$ and we use similar notations in this paper.  The boundary wavefunctions formally read:
\begin{subequations}
\label{eq:PsiS}
\begin{align}
    &\Psi_{\mathrm{S}}(\phi_+(+\infty)) = \langle \phi_+(+\infty), \phi_+^\ast(+\infty); +\infty| \hatSS^\dagger \outket \,,\\
    &\Psi_{\mathrm{S}}^\ast(\phi_-(+\infty)) = \outbra \hatSS |\phi_-(+\infty), \phi_-^\ast(+\infty); +\infty\rangle \,.
\end{align}
\end{subequations}
These explicit forms will be found in Sec.~\ref{sec:boundary}.
We note that $J$ and $J^\ast$ in the generating functional represent $J_+$, $J_+^\ast$, $J_-$, and $J_-^\ast$ collectively.  The action $S$ corresponds to the Lagrangian density in Eq.~\eqref{eq:Lag_scalar}.  As emphasized before, it is important to keep in mind that $\mathcal{C}_+$ extends backward from $x^0=+\infty$ to $x^0=-\infty$, and $\mathcal{C}_-$ corresponds to the forward evolution in time.  The last delta function imposes the constraint of connecting $\mathcal{C}_\pm$ at $x^0=-\infty$.

%%%%%
\paragraph{QED:}

One can immediately write down the counterpart for the Dirac field as
\begin{align}
    Z[\eta,\bar\eta] &= \int [\rmd\psi_+][\rmd\bar{\psi}_+]\; \Psi_{\mathrm{D}}(\psi_+(+\infty))\; \exp\Big[ -\rmi S(\psi_+,\bar{\psi}_+) - \rmi\int \rmd^4x\,(\bar{\eta}_+ \psi_+ + \bar{\psi}_+ \eta_+) \Big] \times \notag\\
    &\quad\times \int[\rmd\psi_-][\rmd\bar{\psi}_-]\; \Psi_{\mathrm{D}}^\ast(\psi_-(+\infty))\; \exp\Big[ +\rmi S(\psi_-,\bar{\psi}_-) + \rmi\int \rmd^4x\,(\bar{\eta}_- \psi_- + \bar{\psi}_- \eta_-) \Big] \times \notag\\ 
    &\quad\times \delta\big[\psi_+(-\infty) - \psi_-(-\infty)\big] \, \delta\big[\bar{\psi}_+(-\infty) - \bar{\psi}_-(-\infty)\big] \,,
    \label{eq:PathInt_Dirac}
\end{align}
where the boundary wavefunctions similarly read:
\begin{subequations}
\label{eq:PsiD}
\begin{align}
    &\Psi_{\mathrm{D}}(\psi_+(+\infty)) = \langle \psi_+(+\infty),\bar{\psi}_+(+\infty);+\infty| \hatSD^\dag |0;\text{out}\rangle \,,\\
    &\Psi_{\mathrm{D}}^\ast(\psi_-(+\infty)) = \langle0;\text{out}| \hatSD |\psi_-(+\infty),\bar{\psi}_-(+\infty);+\infty\rangle \,.
\end{align}
\end{subequations}
Here again, the action $S$ on the exponential corresponds to the Lagrangian density in Eq.~\eqref{eq:Lag_dirac}.

%%%%%%%%%%
\subsection{Explicit forms of the boundary wavefunctions}
\label{sec:boundary}

It is essential to incorporate the wavefunctions into the functional integration.  To this end, we should cope with their explicit forms.
As a warm-up exercise, let us first consider a harmonic oscillator problem, which can easily be generalized to the treatment of the annihilation and creation operators for fields.
In the following analysis of a harmonic oscillator, we assume all the quantities to be dimensionless for simplicity. 
We can then compute the wavefunction [see Eq.~\eqref{eq:hatSD_exp}] explicitly from
\begin{align}
    \langle z, z^*|\, \rme^{\sigma \hat{a}^\dagger \hat{b}^\dagger}\, |0\rangle\,.
    \label{eq:squeezed-wave-func}
\end{align}
Here, $\sigma$ is an arbitrary complex parameter.  We define the creation operators, $\hat{a}^\dag$ and $\hat{b}^\dag$, using a complex coordinate, ${z}={x}+\rmi {y}$, and a time-dependent mode function, $f(t)$, as
\begin{subequations}
\begin{align}
    &\langle z,{z}^\ast|\, \hat{a}^\dag
    := -\rmi\Bigl(-\rmi f \frac{\partial}{\partial z} - \dot{f} z^\ast \Bigr)\langle z,{z}^\ast|
    =-\,f\, \rme^{\rmi(\dot{f}/f)|z|^2} \frac{\partial}{\partial z} \rme^{-\rmi(\dot{f}/f)|z|^2} \langle z,{z}^\ast|\,, \\
    &\langle z,{z}^\ast|\, \hat{b}^\dag
    := -\rmi \Bigl(-\rmi f \frac{\partial}{\partial z^\ast} - \dot{f} z \Bigr)\langle z,{z}^\ast| = -\, f\, \rme^{\rmi(\dot{f}/f)|z|^2} \frac{\partial}{\partial  z^\ast} \rme^{-\rmi(\dot{f}/f)|z|^2} \langle z,{z}^\ast|\,.
\end{align}
\end{subequations}
It is easy to confirm that $[\hat{a}^\dag,\, \hat{b}^\dag]=0$ holds.
From these definitions, we can write Eq.~\eqref{eq:squeezed-wave-func} as
\begin{align}
    \langle z, z^\ast|\, \rme^{\sigma \hat{a}^\dagger \hat{b}^\dagger}\, |0\rangle \;\propto\;
    \rme^{\rmi(\dot{f}/f)|z|^2} \, 
    F(\sigma;z,z^\ast)
    := \rme^{\rmi(\dot{f}/f)|z|^2} \, 
    \rme^{f^2\sigma  \frac{\partial}{\partial z}\frac{\partial}{\partial z^\ast}}\,\rme^{-2\rmi(\dot{f}/f)|z|^2}
    \label{eq:squeezed-wave-func2}
\end{align}
using the ground-state wavefunction, i.e., $\langle z,z^\ast|0\rangle \propto \rme^{-\rmi(\dot{f}/f)|z|^2}$.
We can solve $F(\sigma;z,z^\ast)$ as a solution of the following differential equation,
\begin{align}
    \Big( \frac{\rmd}{\rmd\sigma} -f^2 \frac{\partial}{\partial z}\frac{\partial}{\partial {z}^\ast} \Big)\, F(\sigma;z,z^\ast) = 0
\end{align}
with the initial condition,
$F(0;z,z^\ast)=\rme^{-2\rmi (\dot{f}/{f}){|z|^2}}$.  The solution is found to be
\begin{align}
    F(\sigma;z,z^\ast) = \frac{\rme^{-2\rmi(\dot{f}/f)|z|^2}}{1+2\rmi \sigma f\dot{f}}\, \exp\biggl( {\rmi \frac{\dot{f}}{f}\frac{4\rmi \sigma f\dot{f}}{1+2\rmi \sigma f\dot{f}}|z|^2} \biggr) \,,
    \label{eq:F_lam}
\end{align}
leading to the wavefunction of the complex harmonic oscillator as
\begin{align}
    \langle z, z^\ast|\, \rme^{\sigma \hat{a}^\dag \hat{b}^\dag}\, |0\rangle \;\propto\; \rme^{-\rmi (\dot{f}/f)|z|^2}\,\exp\biggl( {\rmi \frac{\dot{f}}{f}\frac{4\rmi\sigma f\dot{f}}{1+2\rmi \sigma f\dot{f}}|z|^2} \biggr)
    \label{eq:harmonic_wavefunc}
\end{align}
up to an overall constant. 
If we normalize the mode function as $f(t)= \rme^{-\rmi \Omega t}/\sqrt{2\Omega}$ with a (dimensionless) frequency $\Omega$, we find the following relation: $2\rmi f\dot{f} = f/f^*$, from which we immediately derive $2\rmi(\dot{f}/{f}) = 1/|f|^2$.  We demand that mode functions under external fields should satisfy this relation (see Appendix~\ref{app:mode-function}).  Then, we can further simplify Eq.~\eqref{eq:harmonic_wavefunc} as  
\begin{align}
    \langle z, z^\ast|\, \rme^{\sigma \hat{a}^\dag \hat{b}^\dag}\, |0\rangle
    \;\propto\;
    \rme^{-\rmi (\dot{f}/f)|z|^2}\,
    \exp\biggl( {\frac{1}{|f|^2}\frac{\sigma\,f/f^\ast}{1+\sigma\, f/f^\ast}|z|^2} \biggr) \,.
    \label{eq:harmonic_wavefunc_final}
\end{align}

%%%
\paragraph{sQED:}

For the complex scalar field, the operator $\hatSS$ similar to Eq.~\eqref{eq:hatSD_exp} takes the form of
\begin{align}
    \hatSS^\dag = \exp\biggl( \int_\bp \, \sigmap^\ast \, \hataout^\dag \, \hatbout^\dag \biggr) \,,
    \qquad
    \sigmap := \frac{\betap}{\alp} \,.
    \label{eq:hatSS_exp}
\end{align}
To rewrite the generating functional in Eq.~\eqref{eq:PathInt_scalar} into a more convenient form, we should express the boundary wavefunction, $\Psi_{\mathrm{S}}$, in terms of the fields.  Specifically, we can rewrite the creation and annihilation operators in terms of the fields in the asymptotic states at $x^0=\pm\infty$ as
\begin{align}
    \hat{a}_\bp^{\text{as}\,\dag}
    &= -\rmi\int_\bx \Big[\, \Upas{p}(x) \, \hat{\pi}(x) - \partial_0 \Upas{p}(x) \, \hat{\phi}^\dag(x)\, \Big] \notag\\
    &= -\rmi \int_\bp \, \rme^{\rmi\bp\cdot\bx} \, \Bigl[\, f_\bp^\text{as}(x^0) \, \hat{\pi}(x) - \dot{f}_\bp^\text{as}(x^0) \, \hat{\phi}^\dag(x)\, \Bigr]_{x^0=\pm\infty} \,,
    \label{eq:a_as}\\
    \hat{b}_{-\bp}^{\text{as}\,\dag}
    &= -\rmi\int_\bx \, \Big[\, \Vpas{p}(x)^\ast \, \hat{\pi}^\dag(x) - \partial_0 \Vpas{p}(x)^\ast \, \hat{\phi}(x)\, \Big] \notag\\
    &= -\rmi\int_\bx \, \rme^{-\rmi\bp\cdot\bx} \Bigl[\, f_\bp^\text{as}(x^0) \, \hat{\pi}^\dag(x) - \dot{f}_\bp^\text{as}(x^0) \, \hat{\phi}(x)\, \Bigr]_{x^0=\pm\infty}\,,
    \label{eq:b_as}
\end{align}
where the explicit forms of $\Upas{p}(x)$, $\Vpas{p}(x)$, and $f_\bp^\mathrm{as}(x^0)$ are given in Appendix~\ref{app:mode-function-sQED}.
Here, the superscript ``as'' indicates the asymptotic states at either $x^0 = -\infty$ (as$\,=\,$in) or $x^0 = +\infty$ (as$\,=\,$out).
Repeating calculation steps similar to the derivation of Eq.~\eqref{eq:harmonic_wavefunc_final}, we can compute the boundary wavefunction $\Psi_{\mathrm{S}}$ in Eq.~\eqref{eq:PsiS} as follows:
\begin{align}
    \Psi_{\mathrm{S}}(\phi_+(+\infty))
    &= \mathcal{N}_{\mathrm{S}}\, \exp\Bigg[\int_{\bx,\by} \, 
    \Bigg\{ \int_\bp \, \rme^{\rmi\bp\cdot(\bx-\by)}
    \frac{\zeta_\bp(\infty)}{|f_\bp^\text{out}(\infty)|^2} \Bigg\}
    \phi_+^\ast(\infty,{\bm x}) \phi_+(\infty,{\bm y})\Bigg] \times \notag\\
    &\quad \times \langle \phi_+(+\infty), \phi_+^\ast(+\infty);+\infty|0;\mathrm{out}\rangle \,,
    \label{eq:PsiS_int}
\end{align}
where $\mathcal{N}_{\mathrm{S}}$ is a normalization constant of the wavefunction and
\begin{equation}
    \zeta_\bp(\infty) :=  \sigmap^\ast
    \frac{f_\bp^\text{out}(\infty)}{f_\bp^\text{out}(\infty)^\ast}
    \Biggl[ 1 + \sigmap^\ast
    \frac{f_\bp^\text{out}(\infty)}{f_\bp^\text{out}(\infty)^\ast} \Biggr]^{-1} \,.
\end{equation}
We also note that the vacuum wavefunction is
\begin{align}
    \langle \phi_+(+\infty), \phi_+^\ast(+\infty);+\infty|0;\mathrm{out}\rangle = 
    \exp\Biggl[ -\rmi\int_{\bx,\by} \, \Biggl\{ \int_\bp \, \rme^{\rmi\bp\cdot(\bx-\by)} \, \frac{\dot{f}_\bp^\mathrm{out}(\infty)}{f_\bp^\mathrm{out}(\infty)} \Biggr\} \phi_+^\ast(\infty,\bx) \phi_+(\infty,\by) \Biggr] \,.
\end{align}
We can obtain $\Psi_{\mathrm{S}}^\ast$ easily by taking the complex conjugate.

%%%
\paragraph{QED:}

The Dirac field is decomposed in terms of the creation and annihilation operators with the spin polarization index as in Eq.~\eqref{eq:vac-transform}, and we already wrote the explicit form of $\hatSD^\dag$ in Eq.~\eqref{eq:hatSD_exp}.
For the Dirac field, the relation between the fields and the annihilation operators is even simpler as
\begin{align}
    \hataass = \int_\bx \, \Upsas{p}{s}(x)^\dag \, \psi(x) \,,
    \qquad
    \hatbass = \int_\bx \, \psi^\dag(x) \, \Vpsas{p}{s}(x) \,,
\end{align}
where the explicit forms of $\Upsas{p}{s}(x)$ and $\Vpsas{p}{s}(x)$ are given in Appendix~\ref{app:mode-function-QED}.
Therefore, we can express the boundary wavefunction as
\begin{align}
  \Psi_{\mathrm{D}}(\psi_+(+\infty))
  &= \calN_{\mathrm{D}} \exp\Bigg[\int_{\bx,\by} \, \bar{\psi}(\infty,\bx) \Bigg\{ \sum_s\int_\bp \, \sigmap^\ast\, \gamma^0\, \Upsout{p}{s}(x) \, \Vpsout{p}{s}(y)^\dag \Bigg\} \psi(\infty,\by) \Bigg] \times \notag\\
  &\quad\times \langle{\psi}_+(+\infty)\bar{\psi}_+(+\infty);+\infty|0;{\rm out}\rangle \,.
  \label{eq:PsiD_int}
\end{align}

%%%%%%%%%%
\subsection{Combined expressions of the generating functionals}

We plug Eq.~\eqref{eq:PsiS_int} into Eq.~\eqref{eq:PathInt_scalar} for sQED and Eq.~\eqref{eq:PsiD_int} into Eq.~\eqref{eq:PathInt_Dirac} for QED, respectively.
Then, we obtain the full functional integration representations of the generating functionals.
Here, we note that our expression in Eq.~\eqref{eq:PsiS_int} is decomposed into two pieces with and without $\zeta_\bp(\infty)$.  The exponential factor without $\zeta_\bp(\infty)$ is unrelated to the Bogoliubov transformation, simply representing the free Fock vacuum.
This factor ensures the convergence of the functional integration at the boundary and provides the $\rmi\epsilon$ term in the action~\cite{Weinberg:1995mt,Weinberg:2005vy}. 
Similarly for the Dirac field, the terms independent of $\sigmap$ represent the free Fock vacuum, and this part serves only to provide the $\rmi\epsilon$ term. 
Keeping this in mind, we can make further simplification.

%%%
\paragraph{sQED:}

We rewrite Eq.~\eqref{eq:PathInt_scalar} using an index, $a=\pm$, into
\begin{align}
    Z[J,J^*] = \int_{C_{-\infty}} \!\! [\rmd\phi_a][\rmd\phi_a^\ast] \, \exp\biggl[ -\rmi S_+(\phi_+, \phi_+^*) + \rmi S_-(\phi_-,\phi_-^*) + \rmi\int \rmd^4x\, c^{ab} (J_a \phi_b + J_a^* \phi_b^*) \biggr]\,.
    \label{eq:PathInt_scalar2}
\end{align}
We omit the overall normalization factor, for it is irrelevant for computing expectation values.
Also, we set the metric for the sum over $a$ to $c^{++}=1$, $c^{--}=-1$, $c^{+-}=c^{-+}=0$.
The integration contour is constrained by $C_{-\infty}$ which indicates the conditions of $\phi_+(-\infty) = \phi_-(-\infty)$ and $\phi_+^\ast(-\infty) = \phi_-^\ast(-\infty)$ as imposed in Eq.~\eqref{eq:PathInt_scalar}.

Furthermore, in the above expression, we introduced $S_\pm$ along the closed-time path, incorporating the boundary wavefunctions, which are defined as
\begin{subequations}
\begin{align}
    S_+(\phi,\phi^*) &:= S_{-\rmi \epsilon} + \rmi \int \rmd^4x\,\rmd^4y \, \bar\Sigma_{\mathrm{S}}(x,y)\,\phi^*(x)\phi(y) \,,\\
    S_-(\phi,\phi^*) &:= S_{+\rmi \epsilon} - \rmi\int \rmd^4x\,\rmd^4y \, \Sigma_{\mathrm{S}}(x,y)\,\phi^*(x)\phi(y) \,,
\end{align}
\end{subequations}
where $S_{\pm\rmi\epsilon}$ denotes the scalar field action in which $m$ is replaced with $m \mp \rmi\epsilon$ in the $\rmi\epsilon$ prescription.  The additional parts from the boundary wavefunctions can be transformed into the form of the self-energy terms which are immediately deduced from Eq.~\eqref{eq:PsiS_int} as
\begin{subequations}
\begin{align}
    & \bar\Sigma_{\mathrm{S}}(x,y) = \int_\bp \, \rme^{\rmi\bp\cdot(\bx-\by)} \frac{\zeta_\bp(\infty)}{|f_\bp^{\mathrm{out}}(\infty)|^2} \, \delta(x^0-\infty)\, \delta(y^0-\infty) \,,\\
    & \Sigma_{\mathrm{S}}(x,y) = \bar\Sigma_{\mathrm{S}}(y,x)^\ast \,.
\end{align}
\label{eq:Sigma_S}
\end{subequations}
We would emphasize that the above expressions hold for any interaction, $V_{\mathrm{S}}(\phi,\phi^\dag)$.

%%%
\paragraph{QED}

For the Dirac field, the generating functional reads:
\begin{align}
     Z[\eta_a,\bar{\eta}_a] = \int_{C_{-\infty}} \!\! [\rmd\psi_a][\rmd\bar{\psi}_a] \, \exp\Big[-\rmi S_+(\psi_+,\bar{\psi}_+) + \rmi S_-(\psi_-,\bar{\psi}_-) + \rmi\int \rmd^4x\, c^{ab}\big(\bar\eta_a \psi_b + \bar{\psi}_a \eta_b\big)\Big]\,.
     \label{eq:PathInt_Dirac2}
\end{align}
The actions, $S_\pm$, along the closed-time path for the Dirac field take the following forms:
\begin{subequations}
\begin{align}
    S_+(\psi,\bar{\psi}) &:= S_{-\rmi\epsilon} + \rmi\int \rmd^4x\, \rmd^4y\; \bar{\psi}(x) \, \bar{\Sigma}_{\mathrm{D}}(x,y) \, \psi(y) \,, \\
    S_-(\psi,\bar{\psi}) &:= S_{+\rmi\epsilon} - \rmi\int \rmd^4x\, \rmd^4y\; \bar{\psi}(x) \, \Sigma_{\mathrm{D}}(x,y) \, \psi(y) \,,
\end{align}
\end{subequations}
where the additional parts, in analogy to the self-energy terms, are given by
\begin{subequations}
\label{eq:Sigma_D}
\begin{align}
    &\bar{\Sigma}_{\mathrm{D}}(x,y) := \sum_s \int_\bp \, \sigmap^\ast \,\gamma^0\, \Upsout{p}{s}(x) \, \Vpsout{p}{s}(y)^\dag \, \delta(x^0-\infty) \, \delta(y^0-\infty) \,,\\
    &{\Sigma}_{\mathrm{D}}(x,y) := \bar{\Sigma}_{\mathrm{D}}(y,x)^\dag \,.
%    &{\Sigma}_{\mathrm{D}}(x,y) := \sum_s \int_\bp \, \sigmap\,\gamma^0\, \Vpsout{p}{s}(x) \, \Upsout{p}{s}(y)^\dag \, \delta(x^0-\infty) \, \delta(y^0-\infty)\,.
\end{align}
\end{subequations}
Again, the above expressions are intact in the presence of the interaction, so that only the propagators are modified by the boundary wavefunctions.

%%%%%%%%%%%%%%%%%%%%
\section{Propagators}
\label{sec:propagator}
%%%%%%%%%%%%%%%%%%%%

In principle, we can compute any physical observables from the generating functionals in Eqs.~\eqref{eq:PathInt_scalar2} and \eqref{eq:PathInt_Dirac2}, but in practice, to formulate the perturbation theory, we need to find the modified propagators.

%%%%%%%%%%
\subsection{Formal relations of the real-time propagators}

Because of the constraints at the turning point of the closed-time path, the correlations between fields on $\mathcal{C}_+$ and $\mathcal{C}_-$ remain; hence, the propagators have four components.

%%%
\paragraph{sQED:}
From the generating functionals in Eqs.~\eqref{eq:PathInt_scalar2} and \eqref{eq:PathInt_Dirac2}, we can read out the full propagators including $\Sigma$ and $\bar\Sigma$ for the complex scalar fields.

Let us first define the ordinary time-ordered and anti-time-ordered propagators as
\begin{subequations}
\label{eq:D0-1}
\begin{align}
    &D_0^{--}(x,y) := \rmi (-D^2_x - m^2 + \rmi\epsilon)^{-1} \, \delta^{(4)}(x-y) \,,\\
    &D_0^{++}(x,y) := -\rmi (-D^2_x - m^2 - \rmi\epsilon)^{-1} \, \delta^{(4)}(x-y) \,.
\end{align}
\end{subequations}
It should be noted that these are not the free propagators.  We use the subscript ``0'' to indicate the propagators without the effects of the boundary wavefunctions, in the same sense as $S_0(x,y)$ in Eq.~\eqref{eq:S}.  Including the boundary wavefunctions in a form of the self energy, we can write down the $2\times 2$ real-time propagators, among which two are
\begin{subequations}
\label{eq:D0-2}
\begin{align}
    & D^{--}(x,y)^{-1} = D_0^{--}(x,y)^{-1} - \Sigma_\mathrm{S}(x,y) \,,
    \label{eq:D--} \\
    & D^{++}(x,y)^{-1} = D_0^{++}(x,y)^{-1} - \bar{\Sigma}_\mathrm{S}(x,y) \,.
\end{align}
\end{subequations}
These are the Dyson equations and we need to take the inversion to get $D^{--}(x,y)$ and $D^{++}(x,y)$.

%%%
\paragraph{QED:}

Similarly, for the Dirac field, we can write down the $2\times 2$ real-time propagators.  The diagonal parts without the boundary wavefunctions are defined by
\begin{subequations}
\begin{align}
    & S_0^{--}(x,y) = -\rmi (\rmi\slashed{D}_x - m + \rmi\epsilon)^{-1} \delta^{(4)}(x-y) \,,\\
    & S_0^{++}(x,y) = \rmi (\rmi\slashed{D}_x - m - \rmi\epsilon)^{-1} \delta^{(4)}(x-y) \,.
\end{align}
\label{eq:S0}
\end{subequations}
Then, including the boundary wavefunctions, the propagators take the following form:
\begin{subequations}
\begin{align}
    & S^{--}(x,y)^{-1} = S_0^{--}(x,y)^{-1} + \Sigma_{\mathrm{D}}(x,y)\,,
    \label{eq:S--} \\
    & S^{++}(x,y)^{-1} = S_0^{++}(x,y)^{-1} + \bar\Sigma_{\mathrm{D}}(x,y)\,.
%    & S^{-+}(x,y) = S^{>}(x,y)\,,\\
%    & S^{+-}(x,y) = -S^{<}(x,y)
\end{align}
\end{subequations}

%%%%%%%%%%
\subsection{Outline of deriving the propagators}
\label{sec:outline}

The remaining task is to perform the self-energy resummation.  Here, we outline the key steps only.

%%%
\paragraph{sQED:}

We solve Eqs.~\eqref{eq:D0-1} and \eqref{eq:D0-2} using the mode functions summarized in Appendix~\ref{app:mode-function}, and obtain the following $2\times 2$ propagators:
\begin{subequations}
\begin{align}
    D^{--}(x,y) &= D_0^{--}(x,y) + \Da(x,y) \,,
    \label{eq:D--Da}\\
    D^{++}(x,y) &= D^{--}(y,x)^\ast \,,
    \label{eq:D++--}\\
    D^{-+}(x,y) &= D^{>}(x,y) = D_0^{>}(x,y) + \Da(x,y) \,,\\
    D^{+-}(x,y) &= D^{<}(x,y) = D_0^{<}(x,y) + \Da(x,y)
\end{align}
\end{subequations}
under the constraints of
\begin{subequations}
\begin{align}
    &\lim_{x_0\to-\infty} \bigl[ D^{--}(x,y)-D^{+-}(x,y) \bigr]=0 \,,\\
    &\lim_{y_0\to-\infty} \bigl[ D^{--}(x,y)-D^{-+}(x,y) \bigr]=0 \,,
\end{align}
\end{subequations}
according to the closed-time path constraint, $C_{-\infty}$.  It is clear that $D^{-+}(x,y)$ and $D^{+-}(x,y)$ are then consistent with the following decomposition in terms of $D^>(x,y)$ and $D^<(x,y)$, that is,
\begin{align}
    D^{--}(x,y) = \theta(x^0-y^0)\,D^{>}(x,y) + \theta(y^0-x^0)\,D^{<}(x,y)\,.
\end{align}

The explicit forms in terms of the mode functions are
\begin{align}
    & D^{>}_0(x,y) = \int_\bp \, \frac{1}{\alp^\ast} \, \Upout{p}(x)\,\Upin{p}(y)^\ast \,,
    \label{eq:D-0def} \\
    & D^{<}_0(x,y) = \int_\bp \, \frac{1}{\alp^\ast} \, \Vpin{p}(x)\,\Vpout{p}(y)^\ast \,, \\
    & \Da(x,y) = \int_\bp \, \frac{\betap}{\alp^\ast} \, \Vpin{p}(x)\,\Upin{p}(y)^\ast \,.
    \label{eq:Dadef}
\end{align}
It is important to note that $D^{--}(x,y)^{-1}$ in Eq.~\eqref{eq:D--} should be converted into $D^{--}(x,y)$ in Eq.~\eqref{eq:D--Da}, and the resummation procedures are encoded in deriving $\Da(x,y)$, which is summarized in Appendix~\ref{app:derivation}.
In what follows, we further analyze mainly Eqs.~\eqref{eq:D-0def}--\eqref{eq:Dadef}.
Using these representations together with the definitions of $\Sigma_{\mathrm{S/D}}$ and $\bar{\Sigma}_{\mathrm{S/D}}$ in Eqs.~\eqref{eq:Sigma_S} and \eqref{eq:Sigma_D}, we can confirm the relation in Eq.~\eqref{eq:D++--}.

By substituting the mode functions in Appendix~\ref{app:mode-function-sQED} into Eqs.~\eqref{eq:D-0def}--\eqref{eq:Dadef}, we obtain
\begin{align}
    & D^{>}_0(x,y) = -\rmi \frac{\rme^{\frac{\rmi 3\pi}{4}}}{\sqrt{4\pi}} \, \rme^{\frac{\rmi eE}{2} z_3 (x_0+y_0)}
    \intperp \, \frac{\rme^{\rmi\bp_\perp \cdot \bz_\perp}}{(2\pi)^3} \, \Gamma(-\nus) \, \mathscr{I}^{(-)}_{\nus}(\bp_\perp;z_0,z_3) \,,
    \label{eq:D0>J} \\
    & D^{<}_0(x,y) = -\rmi \frac{\rme^{\frac{\rmi 3\pi}{4}}}{\sqrt{4\pi}} \, \rme^{\frac{\rmi eE}{2} z_3 (x_0+y_0)}
    \intperp \, \frac{\rme^{\rmi\bp_\perp \cdot \bz_\perp}}{(2\pi)^3} \, \Gamma(-\nus) \, \mathscr{I}^{(+)}_{\nus}(\bp_\perp;z_0,z_3) \,,
    \label{eq:D0<J} \\
    & \Da(x,y) = -\rmi \frac{\rme^{-\frac{\rmi\pi}{4}}}{\sqrt{4\pi}} \, \rme^{\frac{\rmi eE}{2} z_3 (x_0+y_0)}
    \intperp \, \frac{\rme^{\rmi \bp_\perp \cdot \bz_\perp}}{(2\pi)^3} \, \rme^{-\rmi\pi\nus} \, \Gamma(-\nus) \, \mathscr{I}^\mathrm{a}_{\nu_s}(\bp_\perp;z_0,z_3) \,,
    \label{eq:DaJ}
\end{align}
where, for the scalar field, the dimensionless transverse momentum is defined as
\begin{align}
    \nus := -\rmi\frac{(\bp_\perp^2+m^2)}{2eE} - \frac{1}{2} \,.
    \label{eq:nus}
\end{align}
Also, according to Ref.~\cite{Fradkin:1991zq}, the integrals  $\mathscr{I}^{(\pm)}_\nu(\bp_\perp;z_0,z_3)$ and $\mathscr{I}^\mathrm{a}_\nu(\bp_\perp;z_0,z_3)$ are defined as 
\begin{align}
    \mathscr{I}^{(\mp)}_\nu(\bp_\perp;z_0,z_3) &:= \int_{-\infty}^{+\infty} \! \rmd q\, \rme^{\rmi\sqrt{eE}z_3 q} \,
    D_{\nu} \Bigl(\mp(1+\rmi)\Bigl( q-\tfrac{\sqrt{eE}}{2}z_0 \Bigr)\Bigr)
    \, D_{\nu} \Bigl( \pm(1+\rmi)\Bigl( q+\tfrac{\sqrt{eE}}{2}z_0 \Bigr)\Bigr) \,,
    \label{eq:Jmpdef} \\
    \mathscr{I}^{\rm a}_\nu(\bp_\perp;z_0,z_3) &:= \int_{-\infty}^{+\infty} \! \rmd q \, \rme^{\rmi\sqrt{eE}z_3q} \,
    D_{\nu}\Bigl( (1+\rmi)\Bigl( q-\tfrac{\sqrt{eE}}{2}z_0 \Bigr)\Bigr)
    \, D_{\nu} \Bigl( (1+\rmi) \Bigl( q+\tfrac{\sqrt{eE}}{2}z_0 \Bigr)\Bigr) \,.
    \label{eq:Jadef}
\end{align}

%%%
\paragraph{QED:}

In the same way, we can write down the $2\times 2$ propagators for the Dirac field as
\begin{subequations}
\begin{align}
    S^{--}(x,y) &= S_0^{--}(x,y) + \Sa(x,y) \,,
    \label{eq:S--Sa}\\
    S^{++}(x,y) &= \gamma^0 \, [S^{--}(y,x)]^\dag \, \gamma^0 \,,\\
    S^{-+}(x,y) &= S^{>}(x,y) = S_0^{>}(x,y) + \Sa(x,y) \,,\\
    S^{+-}(x,y) &= -S^{<}(x,y) = -S_0^{<}(x,y) + \Sa(x,y)
\end{align}
\end{subequations}
under the constraint of
\begin{subequations}
\begin{align}
    &\lim_{x_0\to -\infty} \bigl[ S^{--}(x,y) - S^{+-}(x,y) \bigr] = 0 \,,\\
    &\lim_{y_0\to -\infty} \bigl[ S^{--}(x,y) - S^{-+}(x,y) \bigr] = 0 \,.
\end{align}
\end{subequations}
which are consistent with
\begin{align}
    S^{--}(x,y) = \theta(x^0-y^0)\,S^{>}(x,y) - \theta(y^0-x^0)\,S^{<}(x,y)\,.
\end{align}

The explicit forms by means of the mode functions are
\begin{align}
    & S^{>}_0(x,y) = -\sum_s \int_\bp \, \frac{1}{\alp^\ast} \, \Upsout{p}{s}(x) \, \bUpsin{p}{s}(y) \,,
    \label{eq:S-0def} \\
    & S^{<}_0(x,y) = -\sum_s \int_\bp \, \frac{1}{\alp^\ast} \, \Vpsin{p}{s}(x) \, \bVpsout{p}{s}(y) \,,\\
    & \Sa(x,y) = \sum_s \int_\bp \, \frac{\betap}{\alp^\ast} \, \Vpsin{p}{s}(x) \, \bUpsin{p}{s}(y) \,,
    \label{eq:Sadef}
\end{align}
which are, as a result of substituting Eqs.~\eqref{eqapp:ups} and \eqref{eqapp:vps} for the mode functions, transformed into
\begin{align}
    S^{>}_0(x,y)
    &= -\rmi (\rmi\slashed{D}_x+m) \Biggl\{\intperp \, \frac{\rme^{\rmi\bp_\perp\cdot\bz_\perp}}{(2\pi)^3} \, \frac{\rme^{-\rmi\frac{\pi}{4}}}{\sqrt{4\pi}} \,
    \Gamma(-\nud) \, \rme^{\rmi\frac{eE}{2}z_3(x_0+y_0)} \times \notag\\
    &\quad\times \biggl[ \frac{1+\gamma^0\gamma^3}{2} \mathscr{I}^{(-)}_{\nud}(\bp_\perp;z_0,z_3) - \nud\frac{1-\gamma^0\gamma^3}{2} \mathscr{I}^{(-)}_{\nud-1}(\bp_\perp;z_0,z_3) \biggr]\Biggr\} \,,
    \label{eq:S0>-J}\\
    S^{<}_0(x,y)
    &= \rmi (\rmi\slashed{D}_x+m) \Biggl\{ \intperp \, \frac{\rme^{\rmi\bp_\perp\cdot\bz_\perp}}{(2\pi)^3} \, \frac{\rme^{-\rmi\frac{\pi}{4}}}{\sqrt{4\pi}} \,
    \Gamma(-\nud) \, \rme^{\rmi\frac{eE}{2}z_3(x_0+y_0)} \times\notag\\
    &\quad\times \biggl[ \frac{1+\gamma^0\gamma^3}{2} \mathscr{I}^{(+)}_{\nud}(\bp_\perp;z_0,z_3) - \nud\frac{1-\gamma^0\gamma^3}{2} \mathscr{I}^{(+)}_{\nud-1}(\bp_\perp;z_0,z_3) \biggr]\Biggr\} \,,\\
    \Sa(x,y)
    &= \rmi (\rmi\slashed{D}_x+m) \Biggl\{\intperp \, \frac{\rme^{\rmi\bp_\perp\cdot\bz_\perp}}{(2\pi)^3} \, \frac{\rme^{-\rmi\frac{\pi}{4}}}{\sqrt{4\pi}} \, \rme^{-\frac{\pi\lambda}{2}} \,
    \Gamma(-\nud) \, \rme^{\rmi\frac{eE}{2}z_3(x_0+y_0)} \times\notag\\
    &\quad\times \biggl[ \frac{1+\gamma^0\gamma^3}{2} \mathscr{I}^a_{\nud}(\bp_\perp;z_0,z_3) + \nud\frac{1-\gamma^0\gamma^3}{2} \mathscr{I}^a_{\nud-1}(\bp_\perp;z_0,z_3) \biggr]\Biggr\} \,,
    \label{eq:SaJ}
\end{align}
where, for the Dirac field, the dimensionless transverse momentum is
\begin{align}
    \nud := -\rmi\frac{\bp_\perp^2 + m^2}{2eE} \,.
\end{align}
The definitions of $\mathscr{I}^{(\pm)}_\nu(\bp_\perp;z_0,z_3)$ and $\mathscr{I}^{\rm a}_\nu(\bp_\perp;z_0,z_3)$ are the same as the scalar field case. 

%%%%%%%%%%
\subsection{Proper-time integrals for the real-time propagators in a constant electric field}

%--- figure ---%
\begin{figure}
    \centering
    \includegraphics[width=0.4\linewidth]{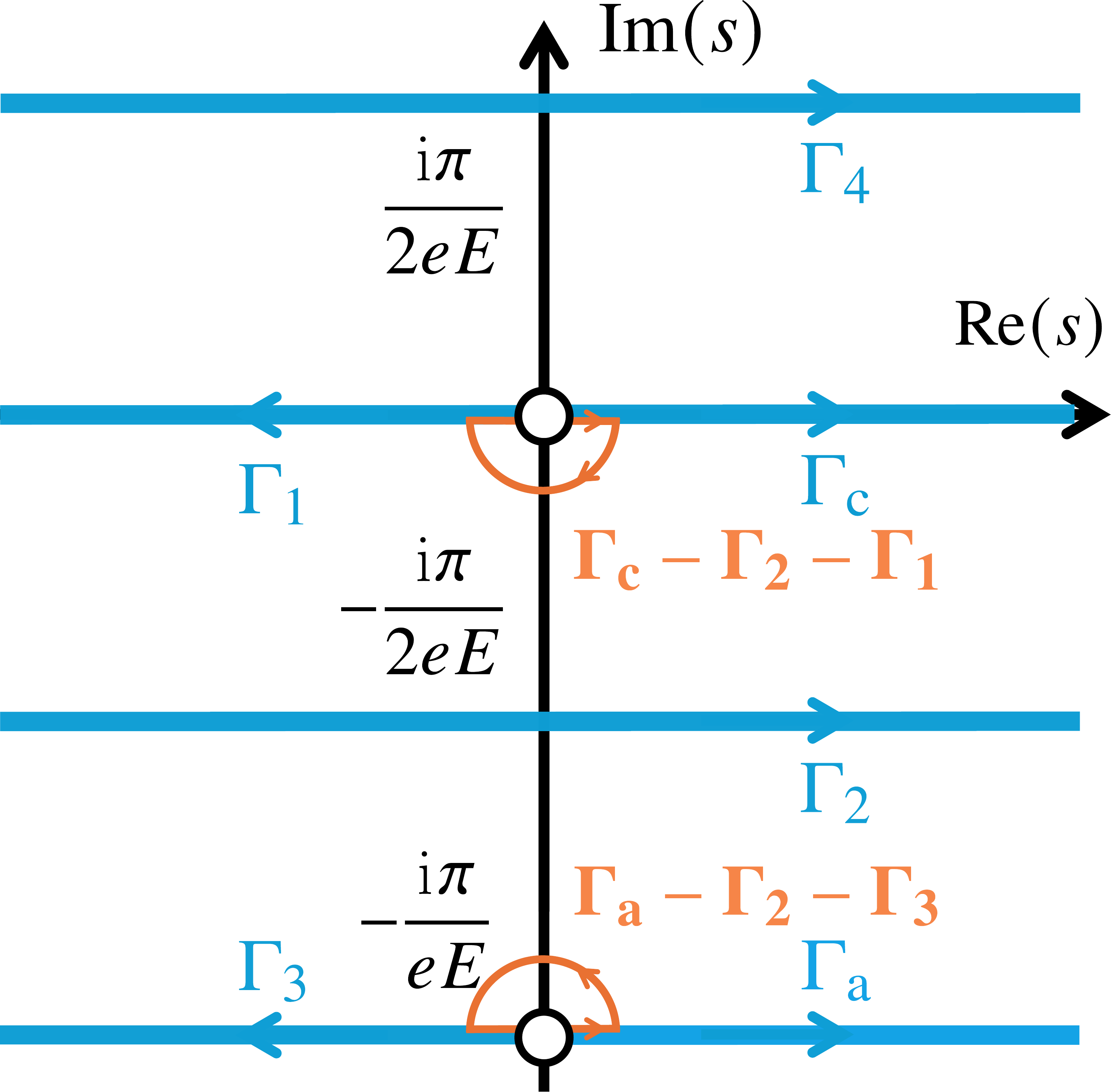}
    \caption{Proper-time integral contours appearing in Eqs.~\eqref{eq:Jmp_prop} and \eqref{eq:Ja_prop} as well as vanishing integrals in Eqs.~\eqref{eq:fds=0at0} and \eqref{eq:fds=0atipi}.}
    \label{fig:path}
\end{figure}
%--- figure ---%

As explained in Appendix~\ref{app:integrals} in details, we can rewrite the integrals in Eqs.~\eqref{eq:Jmpdef} and \eqref{eq:Jadef} in the following proper-time representation:
\begin{align}
    & \mathscr{I}^{(\mp)}_\nu(\bp_\perp;z_0,z_3) = \notag\\
    &= \frac{eE\sqrt{\pi}}{\Gamma(-\nu)} \, \rme^{-\frac{\rmi\pi}{4}} \,
    \biggl\{ \bigl[\theta(\pm z_0) + \theta(\mp z_0)\theta(-\gamma)\bigr] \int\limits_{\Gamma_\mathrm{c}} \rmd s \, I_\nu(s,\gamma)
    + \theta(\mp z_0)\theta(\gamma) \!\! \int\limits_{\Gamma_1+\Gamma_2} \!\! \rmd s\, I_\nu(s,\gamma) \biggr\} \,,
    \label{eq:Jmp_prop} \\
    & \mathscr{I}^{\rm a}_\nu(\bp_\perp;z_0,z_3) = \notag\\
    &= \frac{eE\sqrt{\pi}}{\Gamma(-\nu)} \, \rme^{-\rmi\pi\nu - \rmi\frac{\pi}{4}}
    \biggl\{ \bigl[\theta(-z_3)+\theta(z_3)\theta(\gamma)\bigr] \int\limits_{\Gamma_\mathrm{c}} \rmd s\, I_\nu(s,\gamma)
    + \theta(z_3)\theta(-\gamma) \!\! \int\limits_{\Gamma_1+\Gamma_4} \!\! \rmd s\, I_\nu(s,\gamma) \biggr\} \,,
    \label{eq:Ja_prop}
\end{align}
where $\gamma:=(z_0^2-z_3^2)$ and
\begin{align}
    I_\nu(s,\gamma) := \frac{1}{\sinh(eEs)} \, \exp\biggl[ (2\nu+1)eEs - \rmi\frac{eE}{4}\gamma \coth(eEs) \biggr]\,,
    \label{eq:I-def}
\end{align}
which are valid at least for $-1\leq\Re(\nu)<0$.
The proper-time integration contours, $\Gamma_i$, are shown in Fig.~\ref{fig:path}.
By substituting Eqs.~\eqref{eq:Jmp_prop} and \eqref{eq:Ja_prop} into Eqs.~\eqref{eq:D0>J}--\eqref{eq:DaJ} for scalar field and Eqs.~\eqref{eq:S0>-J}--\eqref{eq:SaJ} for the Dirac field, and performing the remaining momentum integrals, we can obtain the proper-time representations of propagators. 
In the following, we do not give the detailed calculations in the main text but show the final forms only.
See Appendix.~\ref{app:derivation-prop} for the detailed steps.

%--- figure ---%
\begin{figure}
    \centering
    \includegraphics[width=0.4\linewidth]{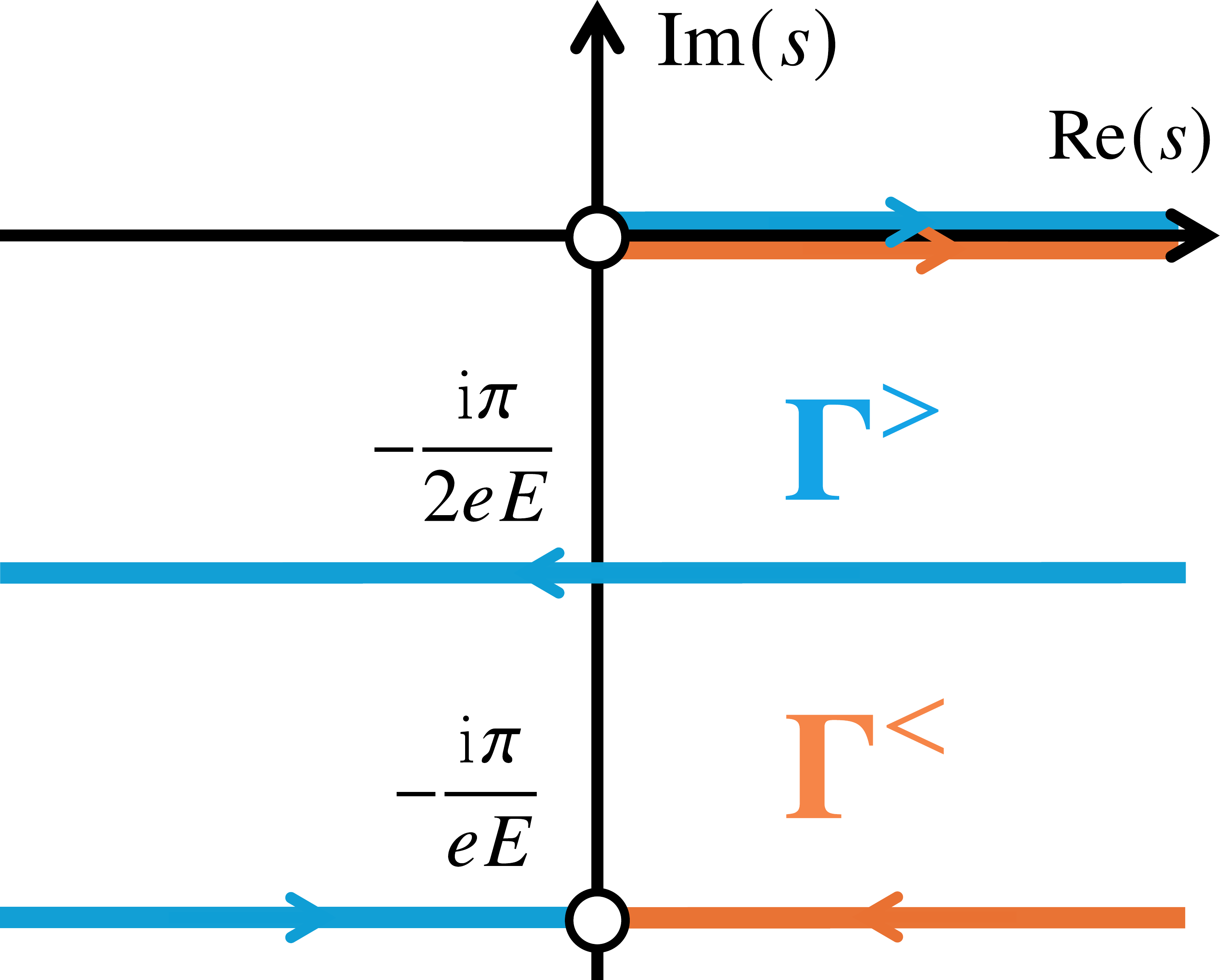}
    \hspace{2em}
    \includegraphics[width=0.4\linewidth]{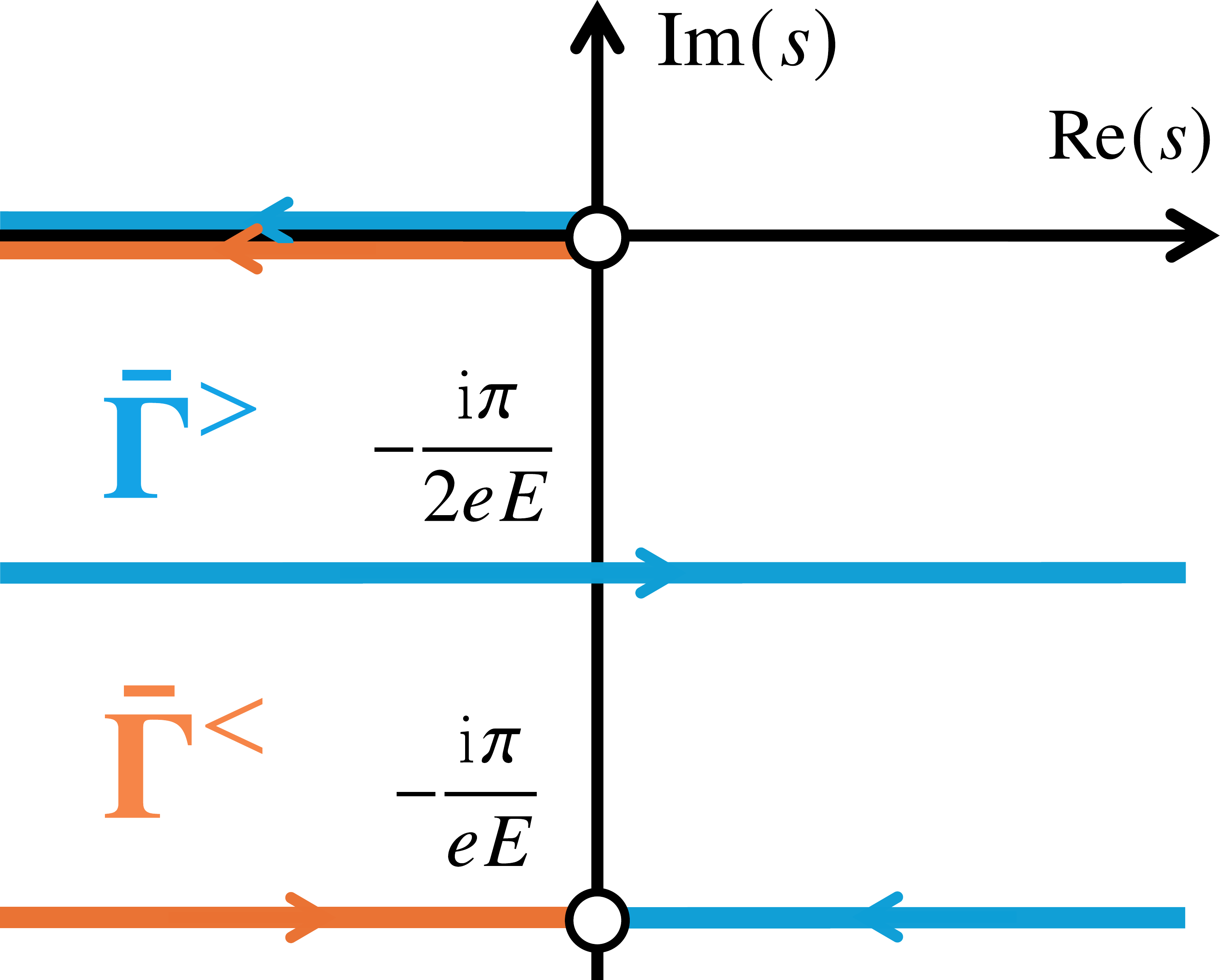}
    \caption{Proper-time integral contours for the $(--)$ and $(++)$ components of the propagators.  The contours along the positive and negative real axis are depicted slightly above and below the real axis just for the presentation purpose; both are simply real integrals over $(0,+\infty)$ and $(-\infty,0)$. Contours passing through (or starting from) $s=-\rmi\pi/(2eE)$ or $s=-\rmi\pi/(eE)$ are parallel to the real axis. The integrals crossing the singularities at $s=0$, $-\rmi\pi/(eE)$ are defined as a principal-value.}
    \label{fig:integ-path-Sc}
\end{figure}
%--- figure ---%

%--- figure ---%
\begin{figure}
    \centering
    \includegraphics[width=0.38\linewidth]{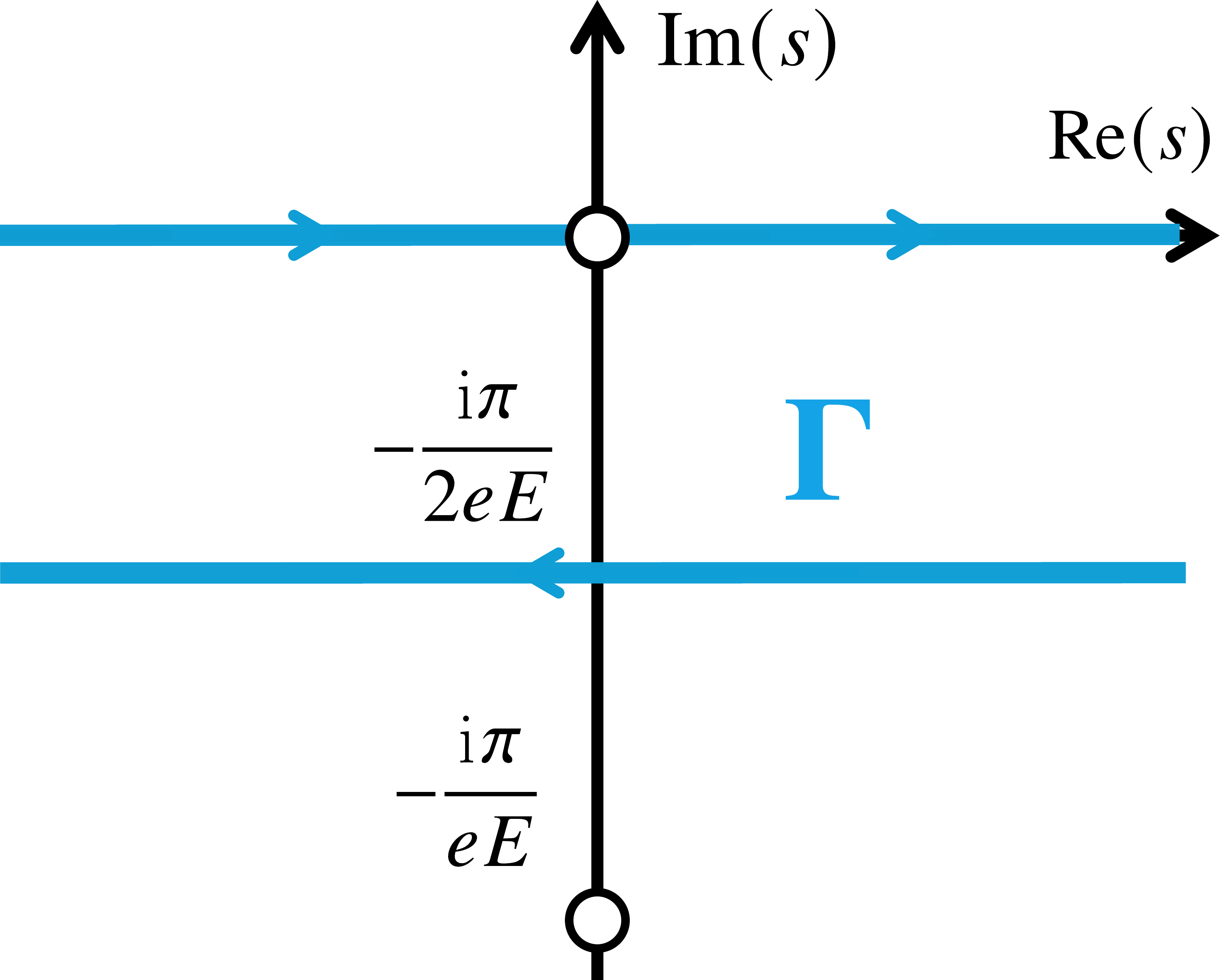}
    \caption{Proper-time integral contours for the $(+-)$ and $(-+)$ components of the propagators.  In the same way as Fig.~\ref{fig:integ-path-Sc}, the contours along the positive and negative real axes are $(0,+\infty)$ and $(-\infty,0)$.}
    \label{fig:integ-path-S}
\end{figure}
%--- figure ---%

%%%
\paragraph{sQED:}

For the complex scalar field, after the calculations in Appendix~\ref{app:proptime-sQED}, we arrive at
\begin{subequations}
\begin{align}
    & D^{--}(x,y) = -\rmi \biggl[ \theta(z_3) \int\limits_{\Gamma^>} \rmd s \, f(s;x,y) + \theta(-z_3) \int\limits_{\Gamma^<} \rmd s \, f(s;x,y) \biggr] \,,
    \label{eq:D--final}\\
    & D^{++}(x,y) = -\rmi \biggl[ \theta(z_3) \int\limits_{\bar{\Gamma}^>} \rmd s \, f(s;x,y) + \theta(-z_3) \int\limits_{\bar{\Gamma}^<} \rmd s \, f(s;x,y) \biggr] \,,
    \label{eq:D++final}\\
    & D^{-+}(x,y) = D^{--}(x,y) - \theta(-z_0) \, \Delta_S(x,y) \,,
    \label{eq:D-+final}\\
    & D^{+-}(x,y) = D^{--}(x,y) - \theta(z_0) \, \Delta_S(x,y) \,,
    \label{eq:D+-final}
\end{align}
\label{eq:D++--final}
\end{subequations}
where a new function,
\begin{align}
    \Delta_S(x,y) := -\rmi\int\limits_\Gamma \rmd s \, f(s;x,y) \,,
\end{align}
is introduced.  The integration contours, $\Gamma^{\gtrless}$ and $\bar{\Gamma}^{\gtrless}$, are shown in Figs.~\ref{fig:integ-path-Sc} and \ref{fig:integ-path-S}.
These full propagators satisfy
\begin{align}
    (D^2_x+m^2) \, D^{\mp \mp}(x,y) = \mp\rmi \delta^{(4)}(x-y) \,,
    \qquad
    (D^2_x+m^2) \, D^{\mp \pm}(x,y) = 0 \,.
\end{align}

%%%
\paragraph{QED:}

Likewise, according to the calculations in Appendix~\ref{app:proptime-QED}, for the Dirac field, the propagators are found to be
\begin{subequations}
\begin{align}
    & S^{--}(x,y) = \rmi(\rmi\slashed{D}_x+m) \biggl[ \theta(z_3) \int\limits_{\Gamma^>} \rmd s \, g(s;x,y) + \theta(-z_3) \int\limits_{\Gamma^<} \rmd s \, g(s;x,y) \biggr] \,,
    \label{eq:S--final}\\
    & S^{++}(x,y) = \rmi(\rmi\slashed{D}_x+m) \biggl[ \theta(z_3) \int\limits_{\bar{\Gamma}^>} \rmd s \, g(s;x,y) + \theta(-z_3) \int\limits_{\bar{\Gamma}^<} \rmd s \, g(s;x,y) \biggr] \,,
    \label{eq:S++final}\\
    & S^{-+}(x,y) = S^{--}(x,y) - \theta(- z_0) \, \Delta_D(x,y) \,,
    \label{eq:S-+final}\\
    & S^{+-}(x,y) = S^{--}(x,y) - \theta( z_0) \, \Delta_D(x,y) \,,
    \label{eq:S+-final}
\end{align}
\label{eq:S++--final}
\end{subequations}
where a new function,
\begin{align}
    \Delta_D(x,y) := \rmi(\rmi\slashed{D}_x+m) \int\limits_\Gamma \rmd s \, g(s;x,y)\,,
\end{align}
is introduced.  They satisfy
\begin{align}
    (\rmi \slashed{D}_x-m) \, S^{\mp \mp}(x,y) = \mp \rmi\delta^{(4)}(x-y) \,,
    \qquad
    (\rmi \slashed{D}_x-m) \, S^{\mp \pm}(x,y) = 0 \,.
\end{align}
For actual calculations, it is important to note that the integrands, $f(s;x,y)$ and $g(s;x,y)$, have singularities at $s=0$ and $s=-\rmi\pi/(eE)$.
The former one at $s=0$ corresponds to the ultraviolet divergence; in fact, $f(s\ll 1/(eE);x,y)$ has no dependence on $eE$ except for the Schwinger phase and the singularity has nothing to do with the electric field.  The latter one at $s=-\rmi\pi/(eE)$, on the other hand, appears from the real-time process associated with the electric field.
One has to treat these singularities in the principal-value prescription~\cite{Schwinger:1960qe, DiPiazza:2018ofz}.

Although these forms of the real-time propagators along the modified proper-time paths are not well-known, they were first derived in canonical quantization several decades ago~\cite{Nikishov:1969tt} and used in some papers such as, e.g., Refs.~\cite{Fradkin:1991zq,Gavrilov:2007hq,Gavrilov:2012jk}. 
In this work, as we stated in the introduction, our aim is to rederive these expressions within the framework of the functional integration formalism, and this aim has now been achieved.  
In this way, it is now clearer which part takes care of resummation and how it is technically performed.  
An interpretation of this resummation in terms of pair production is given in Refs.~\cite{Copinger:2024pai,Copinger:2025ovz}. 
For further technical details that are common to canonical quantization, see Appendix~\ref{app:derivation-prop}.

%%%%%%%%%%%%%%%%%%%%
\section{Applications}
\label{sec:application}
%%%%%%%%%%%%%%%%%%%%

%--- figure ---%
\begin{figure}
    \centering
    \includegraphics[width=0.4\linewidth]{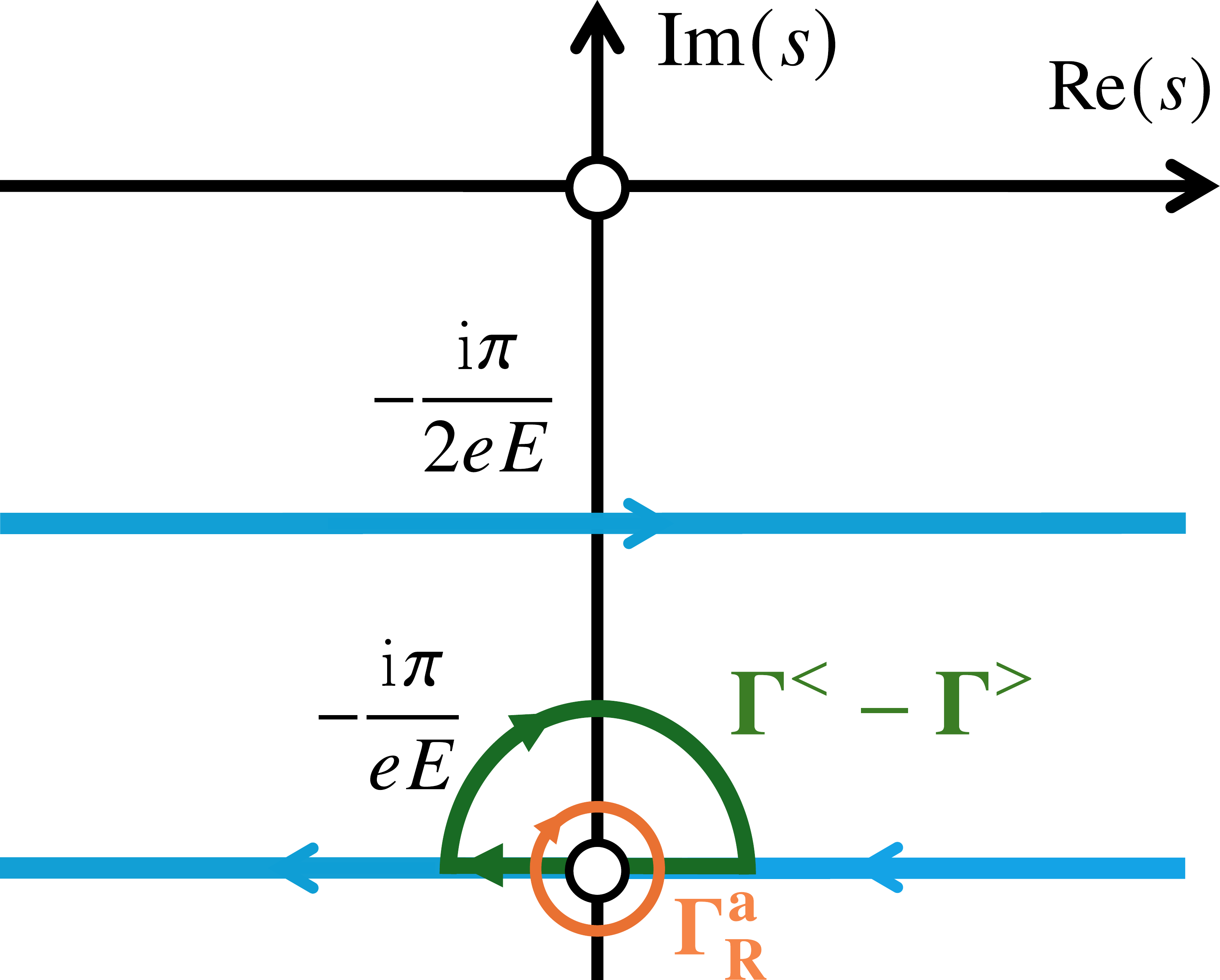}
    \caption{Deformation of the integration contour in Eq.~\eqref{eq:coshf=thetathetaGaRcoshf} from $\Gamma^<-\Gamma^>$ to $\Gamma_{\mathrm{R}}^{\mathrm{a}}$. The pole singularity at $s=-\rmi\pi/(eE)$ on $\Gamma^<-\Gamma^>$ should be understood as the principal-value integral.  In the infinitesimal vicinity of the pole, a lower semicircular contour can harmlessly be added [see discussions around Eqs.~\eqref{eq:gds=0at0} and \eqref{eq:gds=0atipi}], which closes the integration contour around the pole, i.e., deformed into $\Gamma_\mathrm{R}^\mathrm{a}$.}
    \label{fig:integ-path-aR}
\end{figure}
%--- figure ---%

As a concrete example of the application, let us compute the real-time expectation value of the current $j^3(x^0)=\bar{\psi}\gamma^3\psi$ in the one-loop approximation using the formalism described above.
With point-splitting regularization, the current expectation value at one loop reads:
\begin{align}
    \langle j^3(x^0)\rangle
    &= \lim_{x\to y} \mathrm{Tr_s} \bigl[ \gamma^3 S^{--}(x,y) \bigr] \notag\\
    &= \rmi\lim_{x\to y} \mathrm{Tr_s} \biggl\{ \gamma^3 (\rmi\slashed{D}_x+m) \biggl[ \theta(z_3)\int\limits_{\Gamma^>} \rmd s \, g(s;x,y) + \theta(-z_3) \int\limits_{\Gamma^<} \rmd s \, g(s;x,y) \biggr] \biggr\} \,,
\end{align}
where $\mathrm{Tr_s}$ denotes the trace over the spinor indices.  The terms proportional to $m$ vanish after taking the trace, so we can safely neglect them.  Then, by writing $g(s;x,y)$ in terms of $f(s;x,y)$ explicitly, we find
\begin{align}
  \begin{split}
    \langle j^3(x^0)\rangle
    &=-\lim_{x\to y} \mathrm{Tr_s} \biggl\{ \gamma^3 \slashed{D}_x \biggl[ \theta(z_3) \int\limits_{\Gamma^>} \rmd s + \theta(-z_3) \int\limits_{\Gamma^<} \rmd s \biggr] \times \\
    &\qquad\qquad\qquad\times \bigl[ \cosh(eEs)+\sinh(eEs)\gamma^0\gamma^3 \bigr] \, f(s;x,y) \biggr\} \,.
  \end{split}
\end{align}
We can verify that the latter terms in the square brackets proportional to $\sinh(eEs)$ vanish in the limit $x\to y$.  This is because the trace selects $D_0$ and the time derivative of $f(s;x,y)$ is proportional to either $z_3$ or $z_0$.
We thus need to keep only the former terms that contain $\cosh(eEs)$.  We can immediately take the spinor trace to get
\begin{align}
    \langle j^3(x^0)\rangle
    = -4\lim_{x\to y} D_x^3 \biggl\{ \biggl[ \theta(z_3) \int\limits_{\Gamma^>} \rmd s + \theta(-z_3) \int\limits_{\Gamma^<} \rmd s \biggr] \cosh(eEs) \, f(s;x,y) \biggr\} \,.
\end{align}
To proceed further, 
we find it convenient to rewrite the integral contour as follows:
\begin{align}
    \theta(z_3) \int\limits_{\Gamma^>} \rmd s + \theta(-z_3) \int\limits_{\Gamma^<} \rmd s
    = \int\limits_{\Gamma^>} \rmd s + \theta(-z_3)\int\limits_{\Gamma^<-\Gamma^>} \rmd s \,.
\end{align}
In the right-hand side, the contour $\Gamma^>$ in the first term is attached to two singularities at $s=0$ and $s=-\rmi\pi/(eE)$.  The expression of $\langle j^3(x^0)\rangle$ involves $D_x^3$, and we should carefully treat these singularities in taking the derivative in general.  Fortunately, in the present case, these singularities are unrelated to the pair production effect of our interest.  The singularity at $s=0$ corresponds to the ultraviolet divergence, and thus no $eE$-dependent contribution arises.  Besides, in direct calculations, we can show that the singularity at $s=-\rmi\pi/(eE)$ does not generates terms like $\delta(z_3)|_{z_3=0}$ that represents the real-time effect associated with the pair production.
Apart from these singularities, $D_x^3$ directly hits $f(s;x,y)$ in the integrand.  Then, the derivative of the Schwinger phase vanishes in the limit of $x_0\to y_0$, and another derivative is proportional to $z_3$ that goes to zero as  $x_3\to y_3$. 
As a consequence, we conclude that we can safely drop the first term in the following calculations.

In contrast, the second integral along $\Gamma^<-\Gamma^>$ contains the singularity in connection to the pair-production effect.  Because the integration contour can be deformed in the complex plane as long as no singularity is crossed, we can make $\Gamma^< - \Gamma^>$ shrink around $s=-\rmi\pi/(eE)$.
A straightforward calculation shows that within $\Gamma^< - \Gamma^>$, the contribution from the contour segment parallel to the real axis is equivalent to that from a lower semicircular detour around the singularity. 
This enables us to change $\Gamma^< - \Gamma^>$ into $\Gamma_\mathrm{R}^\mathrm{a}$ as illustrated in Fig.~\ref{fig:integ-path-aR}. 
If $z_0^2 - z_3^2 >0$, the integration contribution from the upper infinitesimal semicircle vanishes, and the contribution from a part of the contour parallel to the real axis also disappears in the limit where the radius of the infinitesimal contour tends to zero. 
Thus, the remaining integral is simplified as
\begin{align}
  \begin{split}
    &\int\limits_{\Gamma^<-\Gamma^>} \!\!\! \rmd s \, \cosh(eEs) \, f(s;x,y)
    =
    %\bigl[ \theta(z_0\!-\!0^+)+\theta(-z_0\!-\!0^+) \bigr] \, \theta(z_3^2-z_0^2)
    \theta^+(z_3^2-z_0^2)\int\limits_{\Gamma_\mathrm{R}^\mathrm{a}} \rmd s \, \cosh(eEs)\,f(s;x,y) \,,
  \end{split}
    \label{eq:coshf=thetathetaGaRcoshf}
\end{align}
where $\theta^+(z_3^2-z_0^2):=[\theta(z_0-0^+)+\theta(-z_0-0^+)]\theta(z_3^2-z_0^2)$ to define its value at $z_0=0$.  
We see the necessity of some prescription from the derivative, e.g., $\partial^3 \theta(z_3^2 - z_0^2)
    = 2 z_3 \delta(z_3^2 - z_0^2)
    = 2 \mathrm{sgn}(z_3) \bigl[ \delta(z_3 - z_0) + \delta(z_3 + z_0) \bigr]$.
This expression is subtle; if we perform the $z_3$ integration first and then take the $z_0\to 0$ limit (i.e., $x_0\to y_0$) next, the result should be zero.  
However, if we take the {$z_0\to 0$} limit first, we carefully need to define the sign function, and the result after the integration about $z_3$ is not necessarily zero. 
Thus, this subtlety about the definition of $\theta(z_3^2-z_0^2)$ at $z_0=0$ should be specified, and this form of $\theta^+(z_3^2-z_0^2)$ defines how to treat this subtlety at $z_0=0$. This definition is given in Ref.~\cite{Fradkin:1991zq}. 

Then, the current expectation value becomes
\begin{align}
    \langle j^3(x^0)\rangle
    = -4\lim_{x\to y}\  D_x^3\Big\{ \theta(-z_3)
    \theta^+(z_3^2-z_0^2)\int\limits_{\Gamma_\mathrm{R}^\mathrm{a}} \rmd s \, \cosh(eEs)\,f(s;x,y)\Big\} \,. \label{eq: j3_second}
\end{align}
Now, $D_x^3 f(s;x,y)$ does not have any singularity that survives in the $x\to y$ limit.  Therefore, only a nonzero contribution appears from $D_x^3$ hitting the $z$-dependent coefficient.  In other words, we only need to evaluate the following derivative:
\begin{align}
    \lim_{z\to 0} \partial_x^3 \Bigl\{ \theta(-z_3)\,\theta(z_3^2-z_0^2)\,[\theta(z_0-0^+)+\theta(-z_0-0^+)] \Bigr\} 
    = -\delta(z_3) \Bigr|_{z_3= 0} \,.
%= -\delta(z_3=0) \,.
\end{align}
From the momentum-space representation of the delta function, we see
\begin{equation}
    \delta(z_3) \Bigr|_{z_3= 0} = \int \frac{\rmd p_3}{2\pi}\, \rme^{-\rmi p_3 z_3} \Bigr|_{z_3= 0} = \frac{p_3^\mathrm{max} - p_3^\mathrm{min}}{2\pi} = \Lambda\, x^0 \,, \label{eq: deltaz3}
\end{equation}
which is because of the change in momentum given by the impulse $\propto x^0$  under the electric force.  Thus, the proportionality coefficient should be $\Lambda \propto eE$.  The last necessary manipulation for Eq.~\eqref{eq: j3_second} is the residue calculation, i.e.,
\begin{align}
    \int\limits_{\Gamma_\mathrm{R}^\mathrm{a}} \rmd s \, \cosh(eEs)\,f(s;x,x) = \frac{eE}{8\pi^2} \, \rme^{-\pi m^2/(eE)} \,,
\end{align}
and, finally we reach the physically sensible expression as
\begin{align}
    \langle j^3(x^0) \rangle
    &= \Lambda \frac{eE\,x^0}{4\pi^3} \, \rme^{-\pi m^2/(eE)} \,.
\end{align}
This result qualitatively agrees with Ref.~\cite{Gavrilov:2007hq} with replacement of $T\leftrightarrow x^0$ in the $T\to \infty$ limit keeping the leading term.

To fix $\Lambda$, we consider the following intuitive argument.  It is a straightforward calculation to estimate the number of created particle pairs in unit volume~\cite{Nikishov:1969tt}, leading to
\begin{align}
    N(x^0) = \frac{(eE)^2\,x^0}{4\pi^3}
    \, \rme^{-\pi m^2/(eE)} \,.
\end{align}
The pair should contribute to the current by the unit of $2$, and thus the current expectation value should satisfy the following relation:
\begin{align}
    \langle j^3(x^0) \rangle = 2N(x^0) \,. \label{eq: discrepancyJ3}
\end{align}
This physically intuitive relation immediately leads to $\Lambda=2eE$.  Indeed, as shown explicitly in Ref.~\cite{Fukushima:2009er}, the $p_3$ distribution of the produced particles extends over $\sim 2eE x^0$ for the Sauter potential with the time duration $\sim x^0$.
We note that a more rigorous treatment of Eq.~\eqref{eq: deltaz3} would need further investigations using exactly solvable fields such as the Sauter potential.
We shall leave this point as a future problem.

%%%%%%%%%%%%%%%%%%%%%%%%%%%%%%%%%%%%%%%%%
\section{Conclusion}
\label{sec:conclusion}
%%%%%%%%%%%%%%%%%%%%%%%%%%%%%%%%%%%%%%%%%

We have constructed an alternative description of the in-in (or the closed-time-path or the Schwinger-Keldysh) formalism suitable for the system with a constant electric field.  Since the electric field causes instability (i.e., pair production of particles and antiparticles) in the vacuum, nontrivial boundary wavefunctions at the infinite past/future should be incorporated into the generating functional. 
Choosing the out-state as the reference vacuum in the in-in perturbative expansion, we recast the boundary wavefunctions with quadratic boundary operators built from the Bogoliubov coefficients. 
In the generating functionals, this procedure introduces boundary-localized quadratic kernels, $\Sigma_{\mathrm{S/D}}$ and $\bar\Sigma_{\mathrm{S/D}}$ for the scalar (S) and the Dirac (D) fields.  

Interestingly, the action is modified by only the quadratic part and the interaction terms are intact. 
After all, we have found that the deviation from the standard in-in setup amounts to the self-energy-like terms in the propagators.
Although the proper-time representations of the propagators in the modified formalism have been known in the literature, the derivation and the physical meaning in the canonical operator formalism look obscure.  Our present derivation based on the functional integration clearly shows that the boundary-wavefunction effects are understood as diagrammatic resummation corresponding to the Dyson series.
The resulting propagators perfectly coincide with the expressions from the canonical operator formalism, but our careful treatment of the closed-time path in the functional integration formalism has clarified the origin of the additional proper-time contours.

As a demonstration of application, we have evaluated an interesting one-loop diagram using the modified propagators. 
The calculation no longer requires some diagrammatic resummation because the modified propagators already encompass the resummation to capture vacuum-instability effects such as the Schwinger pair production. 
Thus, this modified approach advocated in Ref.~\cite{Fradkin:1991zq} and reconfirmed by us provides a practical basis for computing the in-in expectation values with vacuum instability.  We are making further progress in the application of this formalism to real-time quantities, and will report the results elsewhere.

\acknowledgments
The authors would like to give special thanks to Patrick~Copinger and Shi~Pu.  In the early stage of this work, the authors communicated with them, and the present work was strongly inspired through discussions with them.
Their approach is based on the worldline formalism (see Ref.~\cite{Copinger:2025ovz}), while our method does not rely on the worldline but employs the closed-time-path formalism.
The authors also thank
Gergely~Endr\H{o}di and
Gergely~Mark\'{o}
for useful discussions.
This work was partially supported by Japan Society for the Promotion of Science
(JSPS) KAKENHI Grant No.\ 22H01216 (K.F.) and No.\ 24KJ0985 (S.M.).

\appendix

%%%%%%%%%%%%%%%%%%%%
\section{Mode functions and Bogoliubov coefficients}
\label{app:mode-function}
%%%%%%%%%%%%%%%%%%%%

The mode functions in the presence of constant electric field are known in the literature, and we list the definitions and the useful relations to clarify our convention. 

%%%%%
\subsection{Weber parabolic cylinder function}
\label{app:Weber}

As we will see later, the concrete expressions of the mode functions contain the Weber parabolic cylinder functions that solve
\begin{align}
    \frac{\rmd^2 D_\nu(z)}{\rmd z^2} + \biggl(\nu + \frac{1}{2} - \frac{z^2}{4} \biggr) D_\nu(z) = 0\,.
\end{align}
The integral representation of $D_\nu(z)$ is known to be
\begin{align}
    D_\nu(z) = \frac{1}{\sqrt{2\pi}} \exp\Bigl[-\rmi\frac{\pi}{2}\nu + \frac{z^2}{4}\Bigr]
    \int_{-\infty}^{+\infty} t^\nu \rme^{-\frac{t^2}{2} + \rmi t z} \,\rmd t
    \label{eq:D-def}
\end{align}
for $\Re(\nu)>-1$.
These functions satisfy the following recurrence relations:
\begin{align}
  \frac{\rmd D_\nu(z)}{\rmd z} = \frac{z}{2} D_\nu(z) - D_{\nu+1}(z) 
  = -\frac{z}{2} D_\nu(z) + \nu D_{\nu-1}(z) \,. 
\end{align}
Also, for practical calculations, the following formulas are useful, that is,
\begin{align}
  & D_\nu(z) = \rme^{\pm \rmi\pi\nu} D_\nu(-z) + \frac{\sqrt{2\pi}}{\Gamma(-\nu)}
    \,\rme^{\pm \frac{\rmi\pi}{2}(\nu+1)} D_{-\nu-1}(\mp \rmi z) \,,\\
  & D_\nu(z) D_{\nu-1}(-z) + D_{\nu-1}(z) D_{\nu}(-z) = \frac{\sqrt{2\pi}}{\Gamma(1-\nu)} \,,\\
  & D_\nu(z) D_{-\nu}(\rmi z) - \rmi\nu D_{\nu-1}(z) D_{-\nu-1}(\rmi z) = \rme^{-\frac{\rmi\pi}{2}\nu} \,,\\
  & D_{\nu}(z) = \frac{\Gamma(\nu+1)}{\sqrt{2\pi}} \Bigl[ \rme^{-\frac{\rmi\pi}{2}\nu}D_{-\nu-1}(-\rmi z) + \rme^{\frac{\rmi\pi}{2}\nu} D_{-\nu-1}(\rmi z) \Bigr] \,. 
\end{align}

%%%%%%%%%%
\subsection{Explicit expressions in sQED}
\label{app:mode-function-sQED}

To simplify lengthy expressions of the mode functions, we introduce the following variables, 
\begin{align}
    \tau := \sqrt{\frac{2}{eE}}\,(p_3-eE x^0) \,,\qquad 
    \lambda := \frac{m^2+\bp_\perp^2}{eE} \,. 
\end{align}
The mode functions are independent solutions of the equation of motion.  Thus, for the complex scalar field in sQED, $\Upas{p}(x)$ and $\Vpas{p}(x)$ are the solutions of
\begin{align}
    \bigl[ \partial_0^2 - \partial_{\perp}^2 - (\partial_3-\rmi eEx^0)^2 - m^2 \bigr] \Phi(x) = 0 \,.
%    &\bigl[ \partial_0^2 - \partial_{\perp}^2 - (\partial_3-\rmi eEx^0)^2 - m^2 \bigr]\Vpas{p}(x) = 0\,.
\end{align}
Below, we indicate the asymptotic states of both ``in'' and ``out'' collectively as ``as'' for simplicity.
These mode functions satisfy orthogonality as
\begin{align}
    &\langle\Upas{p}(x^0,\bx),\Upas{q}(x^0,\bx)\rangle_{\text{KG}} = (2\pi)^3\delta^{(3)}( \bp-\bq)\,,\\
    &\langle\Vpas{p}(x^0,\bx),\Vpas{q}(x^0,\bx)\rangle_{\text{KG}} = (2\pi)^3\delta^{(3)}(\bp-\bq)\,,\\
    &\langle\Upas{p}(x^0,\bx),\Vpas{q}(x^0,\bx)\rangle_{\text{KG}} = 0
\end{align}
with the Klein-Gordon inner product,
\begin{align}
    \langle f(x),g(x)\rangle_{\text{KG}} := \rmi \int_{\bx} \big(f^\ast\partial_{0} g - \partial_{0} f^\ast g \big)\,. 
\end{align}
They also satisfy the completeness relations, i.e.,
\begin{align}
    &\int_{\bp} \Bigl[ \Upas{p}(x^0,\bx) \Upas{p}(x^0,\by)^*-\Vpas{p}(t,{\bm x})\Vpas{p}(t,{\bm y})^*\Big]=0,\\
    &\rmi\int_{\bp} \Big[ \partial_0 \Upas{p}(x^0,\bx) \Upas{p}(x^0,\by)^\ast - \partial_0 \Vpas{p}(x^0,\bx) \Vpas{p}(x^0,\by)^\ast \Bigr] = \delta^{(3)}(\bx-\by) \,. 
\end{align}

For particles and antiparticles, we decompose the temporal dependence in the as=in/out mode functions as
\begin{align}
  \Upas{p}(x) = \rme^{\rmi \bp\cdot \bx} \, f_{\bp}^{\text{as}}(x^0) \,, 
  \qquad 
  \Vpas{p}(x) = \rme^{\rmi \bp\cdot \bx} \, f_{\bp}^{\text{as}}(x^0)^\ast \,. 
  \label{sec:inout-mode}
\end{align}
The functions $f_\bp^{\text{as}}$ are two independent solutions of the Klein-Gordon equation in momentum space.
From the asymptotic boundary conditions, they are given by
\begin{align}
  & f^{\text{in}}_{\bp}(x^0) = \frac{\rme^{-\frac{\pi}{8}\lambda}}{(2eE)^{1/4}} \,
    D_{\frac{\rmi\lambda}{2}-\frac{1}{2}} \bigl(\rme^{-\frac{\rmi\pi}{4}}\tau\bigr)
    \quad \xrightarrow{x^0\to -\infty} \quad
    \frac{\rme^{-\rmi\int^{x^0} \omega_{\bp}(t)\rmd t}}{\sqrt{2\omega_{\bp}(x^0)}} \,,
    \label{eq:in-mode-asym} \\
  & f^{\text{out}}_{\bp}(x^0) = \frac{\rme^{-\frac{\pi}{8}\lambda}}{(2eE)^{1/4}} \,
    D_{-\frac{\rmi\lambda}{2}-\frac{1}{2}} \bigl(-\rme^{\frac{\rmi\pi}{4}}\tau\bigr)
    \quad \xrightarrow{x^0\to +\infty} \quad
    \frac{\rme^{-\rmi\int^{x^0} \omega_{\bp}(t)\rmd t}}{\sqrt{2\omega_{\bp}(x^0)}} \,,
    \label{eq:out-mode-asym}
\end{align}
where the asymptotic forms hold up to an irrelevant phase.  We see that they satisfy
\begin{align}
    \rmi\bigl[ f^{\text{as}}_{\bp}(x^0)^\ast \partial_0 f^{\text{as}}_{\bp}(x^0) - \partial_0 f^{\text{as}}_{\bp}(x^0)^\ast f^{\text{as}}_{\bp}(x^0) \bigr] = 1\,.
\end{align}

The mode functions in the in/out states are related through the Bogoliubov transformation as
\begin{subequations}
\begin{align}
    &\Upin{p}(x) = \alp \Upout{p}(x) + \betap \Vpout{p}(x) \,,\\
    &\Vpin{p}(x) = \betap^\ast \Upout{p}(x) + \alp^\ast \Vpout{p}(x) \,,\\
    &\Upout{p}(x) = \alp^\ast \Upin{p}(x) - \betap \Vpin{p}(x) \,,\\
    &\Vpout{p}(x) = -\betap^\ast \Upin{p}(x) + \alp \Vpin{p}(x) \,.
\end{align}
\end{subequations}
From the asymptotic forms of the Weber parabolic cylinder functions, one can read the Bogoliubov coefficients as
\begin{align}
  \alp = \frac{\sqrt{2\pi}}{\Gamma(\frac{1}{2}-\frac{\rmi\lambda}{2})} \, \rme^{\frac{\rmi\pi}{2}(\frac{\rmi\lambda}{2}+\frac{1}{2})} \,,
  \qquad
  \betap = -\rmi \, \rme^{-\frac{\pi\lambda}{2}} \,.
\end{align}
Using a well-known formula, $\Gamma(1-z)\Gamma(z)=\pi/(\sin(\pi z)$, it is easy to confirm $|\alp|^2-|\betap|^2=1$.

%%%%%%%%%%
\subsection{Explicit expressions in QED}
\label{app:mode-function-QED}

For the Dirac field in QED, the mode functions, $\Upsas{p}{s}(x)$ and $\Vpsas{p}{s}(x)$ are solutions of
\begin{align}
   (\rmi \slashed{D}_x-m)\,\Psi(x) = 0\,,
%  & (\rmi \slashed{D}_x-m)\,\Vpsas{p}{s}(x) = 0\,, 
\end{align}
and satisfy the orthogonality conditions,
\begin{align}
  & \langle\Upsas{p}{s}(x^0,\bx), \Upsas{q}{s'}(x^0, \bx) \rangle_\text{D} = \delta_{ss'}\, (2\pi)^3 \delta^{(3)}(\bp-\bq) \,,\\
  & \langle\Vpsas{p}{s}(x^0,\bx), \Vpsas{q}{s'}(x^0, \bx) \rangle_\text{D} = \delta_{ss'}\, (2\pi)^3 \delta^{(3)}(\bp-\bq) \,,\\
  & \langle\Upsas{p}{s}(x^0,\bx), \Vpsas{q}{s'}(x^0, \bx) \rangle_\text{D} = 0 \,,
\end{align}
and the completeness condition,
\begin{align}
  \sum_s\int_{\bp} \Bigl[ \Upsas{p}{s}(x^0,\bx)\, \Upsas{p}{s}(x^0,\by)^\dag + \Vpsas{p}{s}(x^0,\bx)\,\Vpsas{p}{s}(x^0,\by)^\dag \Bigr] = \,\delta^{(3)}(\bx-\by)
\end{align}
with the inner product of spinors defined by
\begin{align}
    \langle f(x), g(x)\rangle_\text{D} := \int_{\bx} f(x)^\dag g(x) \,.
\end{align}

To obtain the explicit solutions, it is useful to consider the squared Dirac operator:
\begin{align}
  (\rmi \slashed{D} - m) (\rmi \slashed{D} + m)
  = -\bigl[ \partial_0^2 + (p_3- eEx^0)^2 + \bp_\perp^2
  + m^2 + \rmi eE\,\gamma^0\gamma^3 \bigr] \,.
\end{align}
We can classify the solutions by the eigenvalues $\pm1$ of $\gamma^0\gamma^3$.  We shall let $u_1$, $v_1$ belong to the eigenvalue $-1$, and $u_2$, $v_2$ belong to the eigenvalue $+1$.  Their explicit forms in the in/out states are
\begin{align}
  & u_{1 \bp}^{\text{in}}(x^0) = \rme^{-\frac{\pi\lambda}{8}} \, \rme^{\frac{\rmi\pi}{4}} D_{\frac{\rmi\lambda}{2}} \bigl(\rme^{-\frac{\rmi\pi}{4}}\tau\bigr) \,,\\
  & v_{1 \bp}^{\text{in}}(x^0) = \sqrt{\frac{\lambda}{2}} \, \rme^{-\frac{\pi\lambda}{8}} \, \rme^{\frac{\rmi\pi}{4}} D_{-\frac{\rmi\lambda}{2}-1} \bigl(\rme^{\frac{\rmi\pi}{4}}\tau\bigr) \,,\\
  & u_{2 \bp}^{\text{in}}(x^0) = \sqrt{ \frac{\lambda}{2}\frac{\varepsilon_\bp+p_3}{\varepsilon_\bp-p_3} } \,\rme^{-\frac{\pi\lambda}{8}} D_{\frac{\rmi\lambda}{2}-1}\bigl(\rme^{-\frac{\rmi\pi}{4}}\tau\bigr) \,,\\
  & v_{2 \bp}^{\text{in}}(x^0) = -\sqrt{\frac{\varepsilon_\bp+p_3}{\varepsilon_\bp-p_3}}\, \rme^{-\frac{\pi\lambda}{8}}D_{-\frac{\rmi\lambda}{2}} \bigl(\rme^{\frac{\rmi\pi}{4}}\tau\bigr) \,,\\
  & u_{1 \bp}^{\text{out}}(x^0) = \sqrt{\frac{\lambda}{2}} \, \rme^{-\frac{\pi\lambda}{8}} \, \rme^{\frac{\rmi\pi}{4}} D_{-\frac{\rmi\lambda}{2}-1} \bigl(-\rme^{\frac{\rmi\pi}{4}}\tau\bigr) \,,\\
  & v_{1 \bp}^{\text{out}}(x^0) = \rme^{-\frac{\pi\lambda}{8}} \, e^{\frac{\rmi\pi}{4}} D_{\frac{\rmi\lambda}{2}} \bigl(-\rme^{-\frac{\rmi\pi}{4}}\tau\bigr) \,,\\
  & u_{2 \bp}^{\text{out}}(x^0) = \sqrt{\frac{\varepsilon_\bp+p_3}{\varepsilon_\bp-p_3}} \, \rme^{-\frac{\pi\lambda}{8}} D_{-\frac{\rmi\lambda}{2}}\bigl(-\rme^{\frac{\rmi\pi}{4}}\tau\bigr) \,,\\
  & v_{2 \bp}^{\text{out}}(x^0) = -\sqrt{\frac{\lambda}{2}\frac{\varepsilon_\bp+p_3}{\varepsilon_\bp-p_3}} \, \rme^{-\frac{\pi\lambda}{8}} D_{\frac{\rmi\lambda}{2}-1}\bigl(-\rme^{-\frac{\rmi\pi}{4}}\tau\bigr)
\end{align}
with $\varepsilon_\bp^2:=\bp^2 + m^2$.  They satisfy:
\begin{subequations}
\begin{align}
  & \bigl[ \rmi\partial_0 - (p_3-eEx^0) \bigr] u_{1 \bp}^{\text{as}}(x^0) = (\varepsilon_\bp- p_3) u_{2 \bp}^{\text{as}}(x^0) \,,\\
  & \bigl[ \rmi\partial_0 - (p_3-eEx^0) \bigr] v_{1 \bp}^{\text{as}}(x^0) = (\varepsilon_\bp- p_3) v_{2 \bp}^{\text{as}}(x^0) \,,\\
  & \bigl[ \rmi\partial_0 + (p_3-eEx^0) \bigr] u_{2 \bp}^{\text{as}}(x^0) = (\varepsilon_\bp+ p_3) u_{1 \bp}^{\text{as}}(x^0) \,,\\
  & \bigl[ \rmi\partial_0 + (p_3-eEx^0) \bigr] v_{2 \bp}^{\text{as}}(x^0) = (\varepsilon_\bp + p_3) v_{1 \bp}^{\text{as}}(x^0) \,,
\end{align}
\end{subequations}
which are useful to prove the orthogonality and completeness of the mode functions.

Using these solutions, we define the particle/antiparticle mode functions of the Dirac equation by
\begin{align}
  \Upsas{p}{s}(x)
  &:= (\rmi\slashed{D}_x + m) \biggl[ \frac{\rme^{\rmi \bp\cdot \bx}}{\varepsilon_\bp + p_3} \, u_{2\bp}^\text{as}(x_0)\,\xi_s \bigg]
  = \rme^{\rmi \bp\cdot \bx} \biggl[ \gamma^0 u_{1\bp}^\text{as}(x^0) + \frac{\bm{\gamma}^{\perp} \cdot  \bp_\perp + m}{\varepsilon_\bp + p_3} \, u_{2\bp}^\text{as}(x^0) \biggr] \xi_s \,,
  \label{eqapp:ups}\\
  \Vpsas{p}{s}(x)
  &:= (\rmi\slashed{D}_x  + m) \biggl[ \frac{\rme^{\rmi{\bp\cdot \bx}}}{\varepsilon_\bp + p_3}\,v_{2\bp}^\text{as}(x^0) \, \xi_s \biggr]
  = \rme^{\rmi \bp\cdot \bx} \biggl[ \gamma^0 v_{1\bp}^\text{as}(x^0) + \frac{\bm{\gamma}^\perp \cdot  \bp_\perp + m}{\varepsilon_\bp + p_3} \, v_{2\bp}^\text{as}(x^0) \biggr] \xi_s \,.
  \label{eqapp:vps}
\end{align}
Here, the polarizations, $\xi_s$, are chosen from the complete system of $\Sigma^3 = \rmi\gamma^1\gamma^2$ with
\begin{align}
  \gamma^0\gamma^3 \xi_s = \xi_s \,,
  \qquad
  \Sigma^3 \xi_s = s\,\xi_s \,.
\end{align}
One can confirm that they satisfy:
\begin{align}
  \xi_s^\dag\, \xi_{s'} = \delta_{s s'} \,,
  \qquad
  \sum_s \xi_s \, \xi_s^\dag = \frac{1 + \gamma^0 \gamma^3}{2} \,.
\end{align}
One can also confirm the following identities:
\begin{align}
& |u_{1 \bp}^\text{as}(x^0)|^2 + |v_{1 \bp}^\text{as}(x^0)|^2 = 1 \,,\\
& |u_{2 \bp}^\text{as}(x^0)|^2 + |v_{2 \bp}^\text{as}(x^0)|^2 = \frac{\varepsilon_\bp + p_3}{\varepsilon_\bp - p_3} \,,\\
& |u_{1 \bp}^\text{as}(x^0)|^2 + \frac{\varepsilon_\bp - p_3}{\varepsilon_\bp + p_3}|u_{2 \bp}^\text{as}(x^0)|^2 = 1 \,,\\
& |v_{1 \bp}^\text{as}(x^0)|^2 + \frac{\varepsilon_\bp - p_3}{\varepsilon_\bp + p_3}|v_{2 \bp}^\text{as}(x^0)|^2 = 1 \,,\\
& u_{1 \bp}^\text{as}(x^0) u_{2 \bp}^\text{as}(x^0)^\ast = -v_{1 \bp}^\text{as}(x^0) v_{2 \bp}^\text{as}(x^0)^\ast \,,\\
& v_{2 \bp}^\text{in}(x^0) v_{2 \bp}^\text{out}(x^0)^\ast = \frac{\varepsilon_\bp + p_3}{\varepsilon_\bp - p_3} u_{1 \bp}^\text{out}(x^0) u_{1 \bp}^\text{in}(x^0)^\ast \,,\\
& u_{2 \bp}^\text{out}(x^0) u_{2 \bp}^\text{in}(x^0)^\ast = \frac{\varepsilon_\bp + p_3}{\varepsilon_\bp - p_3} v_{1 \bp}^\text{in}(x^0) v_{1 \bp}^\text{out}(x^0)^\ast \,.
\end{align}
The formulae in Appendix~\ref{app:Weber} are useful for proving them.

The mode functions in Eqs.~\eqref{eqapp:ups} and \eqref{eqapp:vps} are related by the Bogoliubov transformations:
\begin{subequations}
\begin{align}
  \Upsin{p}{s}(x) &= \alp \, \Upsout{p}{s}(x) - \betap \, \Vpsout{p}{s}(x) \,,\\
  \Vpsin{p}{s}(x) &= \betap^\ast \, \Upsout{p}{s}(x) + \alp^\ast \, \Vpsout{p}{s}(x) \,,\\
  \Upsout{p}{s}(x) &= \alp^\ast \, \Upsin{p}{s}(x) + \betap \, \Vpsin{p}{s}(x) \,,\\
  \Vpsout{p}{s}(x) &= -\betap^\ast \, \Upsin{p}{s}(x) + \alp \, \Vpsin{p}{s}(x)
\end{align}
\end{subequations}
with the Bogoliubov coefficients given by
\begin{align}
    \alp = \rmi \sqrt{\frac{2}{\lambda}} \, \frac{\sqrt{2\pi}}{\Gamma(-\frac{\rmi\lambda}{2})} \, \rme^{-\frac{\pi\lambda}{4}} \,,
    \qquad
    \betap = -\rme^{-\frac{\pi\lambda}{2}} \,.
\end{align}
It is easy to confirm that they satisfy $|\alp|^2+|\betap|^2=1$.

%%%%%%%%%%%%%%%%%%%%
\section{Explicit calculations of the Dyson series for \texorpdfstring{$D^{--}(x,y)$}{text} and \texorpdfstring{$S^{--}(x,y)$}{text}}
\label{app:derivation}
%%%%%%%%%%%%%%%%%%%%%%%%%%%%%%%%%%%%%%%%%

We show that Eqs.~\eqref{eq:D--Da} and \eqref{eq:S--Sa} hold.
We begin with the scalar field. The $(--)$ component of the full propagator is defined by
\begin{align}
  D^{--}(x,y)=\frac{1}{(D_0^{--})^{-1}-\Sigma_\mathrm{S}} \,,
  \label{eqapp:Dyson-sum-D--}
\end{align}
so we expand the right-hand side as the Dyson series and evaluate each term explicitly.

As a preparation, we define
\begin{align}
  T(x,y) := \int \rmd^4 z \, \Sigma_\mathrm{S}(x,z) \, D_0^{--}(z,y) \,.
\end{align}
A short calculation gives
\begin{align}
  T(x,y) = \int_\bp \, \rme^{\rmi\bp\cdot(\bx-\by)} \, 
  \frac{\zeta_\bp(\infty)^\ast}{\alp^\ast}\,
  \frac{f_\bp^\mathrm{in}(y^0)^\ast}{f_\bp^\mathrm{out}(\infty)^\ast}\,
  \delta(x^0-\infty) \,.
\end{align}
Similarly, the $n$-th product of $T(x,y)$ denoted by
\begin{align}
  T^n(x,y) := \int \rmd^4 x_1 \cdots \rmd^4 x_{n-1} \, T(x,x_1) T(x_1,x_2) \cdots T(x_{n-1},y)
\end{align}
can be written down as
\begin{align}
  T^n(x,y) = \int_\bp \, \rme^{\rmi \bp\cdot(\bx-\by)} \,
  \biggl[\frac{\zeta_\bp(\infty)^\ast}{\alp^\ast}\biggr]^n
  \biggl[\frac{f_\bp^\mathrm{in}(\infty)^\ast}{f_\bp^\mathrm{out}(\infty)^\ast}\biggr]^{n-1}
  \frac{f_\bp^\mathrm{in}(y^0)^\ast}{f_\bp^\mathrm{out}(\infty)^
  \ast} \,
  \delta(x^0-\infty) \,.
\end{align}
Therefore, the $n$-th term in the series for the right-hand side of Eq.~\eqref{eqapp:Dyson-sum-D--} evaluates as
\begin{align}
  \int \rmd^4 z \, D_0^{--}(x,z) \, T^n(z,y) = \int_\bp \, \rme^{\rmi \bp\cdot(\bx-\by)} \, 
  \frac{f_\bp^\mathrm{in}(x^0)^\ast f_\bp^\mathrm{in}(y^0)^\ast}{\alp^\ast} \, 
  \frac{f_\bp^\mathrm{out}(\infty)}{f_\bp^\mathrm{in}(\infty)^\ast}
  \biggl[ \frac{\zeta_\bp(\infty)^\ast}{\alp^\ast}\frac{f_\bp^\mathrm{in}(\infty)^\ast}{f_\bp^\mathrm{out}(\infty)^\ast} \biggr]^n \,.
\end{align}
Using the relation among mode functions,
\begin{align}
  f_\bp^\mathrm{in}(\infty)^\ast = \alp^\ast f_\bp^\mathrm{out}(\infty)^\ast + \betap^\ast f_\bp^\mathrm{out}(\infty) \,,
\end{align}
and a shorthand notation,
\begin{align}
  r = \frac{\betap}{\alp} \, \frac{f_\bp^\mathrm{out}(\infty)^\ast}{f_\bp^\mathrm{out}(\infty)} \,.
\end{align}
We can immediately prove
\begin{align}
  \frac{f_\bp^\mathrm{out}(\infty)}{f_\bp^\mathrm{in}(\infty)^\ast}
  \sum_{n=1}^{\infty} \biggl[ \frac{\zeta_\bp(\infty)^\ast}{\alp^\ast} \frac{f_\bp^\mathrm{in}(\infty)^\ast}{f_\bp^\mathrm{out}(\infty)^\ast} \biggr]^n
  = \frac{1}{\betap^\ast}\frac{r^\ast}{1+r^\ast} \sum_{n=1}^{\infty} \biggl[ \frac{r(1+r^\ast)}{1+r} \biggr]^n
  = \frac{1}{\betap^\ast}\frac{|r|^2}{1-|r|^2}
  = \betap \,.
\end{align}
In the last equality, we use Eq.\eqref{eq:out-mode-asym}.
Then, Eq.~\eqref{eqapp:Dyson-sum-D--} yields
\begin{align}
  D^{--}(x,y) = D^{--}_0(x,y) + \int_\bp \, \frac{\betap}{\alp^\ast} \, \Vpin{p}(x) \, \Upin{p}(y)^\ast \,,
\end{align}
where the last term is $\Da(x,y)$.

We proceed in the same way for the Dirac field. From the definition of the $(--)$ component of full propagator, 
\begin{align}
  S^{--}(x,y) = \frac{1}{(S^{--}_0)^{-1} + \Sigma_\mathrm{D}} \,,
\end{align}
holds.
Introduce, analogously to the scalar case,
\begin{align}
  T_\mathrm{D}(x,y) = \int \rmd^4 z \, \Sigma_\mathrm{D}(x,z) \, S_0^{--}(z,y) \,.
\end{align}
A parallel computation gives
\begin{align}
  \int \rmd^4 z \, S_0^{--}(x,z) \, T_\mathrm{D}^{\;n}(z,y)
  = \sum_s \int_\bp \, \Vpsin{p}{s}(x) \, \bUpsin{p}{s}(y)
  \biggl( \frac{|\betap|^2}{|\alp|^2} \biggr)^n \frac{1}{\alp^\ast \betap^\ast} \,.
\end{align}
Summing the Dyson series, we obtain
\begin{align}
  S^{--}(x,y) = S_0^{--}(x,y) + \sum_s \int_\bp \, \frac{\betap}{\alp^\ast} \,
  \Vpsin{p}{s}(x) \, \bUpsin{p}{s}(y) \,,
\end{align}
where the last term is $\Sa(x,y)$.

%%%%%%%%%%%%%%%%%%%%
\section{Derivation of the proper-time representation}
\label{app:derivation-prop}

We derive the proper-time representations of propagators \eqref{eq:D-0def}--\eqref{eq:Dadef} for the scalar field and \eqref{eq:S-0def}--\eqref{eq:Sadef} for the Dirac field.  
In Sec.~\ref{app:integrals}, we show that Eqs.~\eqref{eq:Jmpdef} and \eqref{eq:Jadef} can be rewritten as Eqs.~\eqref{eq:Jmp_prop} and \eqref{eq:Ja_prop}. 
Then, in Sec.~\ref{app:proptime-sQED}, we derive the proper-time representations of propagators given by \eqref{eq:D++--final} for the scalar field and \eqref{eq:S++--final} for the Dirac field by using the results from Sec.~\ref{app:integrals}.

%%%%%%%%%%
\subsection{ \texorpdfstring{$\mathscr{I}^{(\mp)}_\nu$}{text} and \texorpdfstring{$\mathscr{I}^{\rm a}_\nu$}{text}}
\label{app:integrals}

Below, we outline a derivation of Eq.~\eqref{eq:Jmp_prop} from Eq.~\eqref{eq:Jmpdef} and Eq.~\eqref{eq:Ja_prop} from Eq.~\eqref{eq:Jadef}.  If readers are interested in more details, they can consult the textbook~\cite{Fradkin:1991zq} (containing confusing typos that are corrected in what follows below).  We note that the functions~\eqref{eq:Jmpdef} and \eqref{eq:Jadef} obey
\begin{subequations}
\begin{align}
    & \Bigl( z_0\frac{\partial}{\partial z_3} + z_3\frac{\partial}{\partial z_0} \Bigr) \mathscr{I}^{(\mp)}_\nu(\bp_\perp;z_0,z_3) = 0 \,, \\
    & \Bigl( z_0\frac{\partial}{\partial z_3} + z_3\frac{\partial}{\partial z_0} \Big) \mathscr{I}^\mathrm{a}_\nu(\bp_\perp;z_0,z_3) = 0 \,.
\end{align}
\end{subequations}
This means that $\mathscr{I}^{(\mp)}_\nu$ and $\mathscr{I}^\mathrm{a}_\nu$ are functions of $\gamma$ only, not depending on $z_0$ and $z_3$ separately.  
Therefore, it is convenient to divide the $(z_0,z_3)$-plane into four regions: $z_0>0$ and $\gamma>0$, $z_0<0$ and $\gamma>0$, $z_3>0$ and $\gamma<0$, and $z_3<0$ and $\gamma<0$. 
In each region, we can choose a representative point at which either $z_0$ or $z_3$ vanishes. 
In this way, the functions \eqref{eq:Jmp_prop} and \eqref{eq:Ja_prop} are decomposed as
\begin{align}
    &\mathscr{I}^{(\mp)}_\nu(\bp_\perp;z_0,z_3)
    = \theta(z_0)\theta(\gamma) \mathscr{I}^{(\mp)}_\nu(\bp_\perp;\sqrt{\gamma},0) + \theta(-z_0)\theta(\gamma) \mathscr{I}^{(\mp)}_\nu(\bp_\perp;-\sqrt{\gamma},0) \notag\\
    &\qquad\qquad + \theta(z_3)\theta(-\gamma) \mathscr{I}^{(\mp)}_\nu(\bp_\perp;0,\sqrt{-\gamma}) + \theta(-z_3)\theta(-\gamma) \mathscr{I}^{(\mp)}_\nu(\bp_\perp;0,-\sqrt{-\gamma}) \,,
    \label{eqapp:jmpdecomp} \\
    &\mathscr{I}^\mathrm{a}_\nu(\bp_\perp;z_0,z_3) 
    = \theta(z_0)\theta(\gamma) \mathscr{I}^\mathrm{a}_\nu(\bp_\perp;\sqrt{\gamma},0) + \theta(-z_0)\theta(\gamma) \mathscr{I}^{(\mp)}_\nu(\bp_\perp;-\sqrt{\gamma},0) \notag\\
    &\qquad\qquad +\theta(z_3)\theta(-\gamma) \mathscr{I}^\mathrm{a}_\nu(\bp_\perp;0,\sqrt{-\gamma}) + \theta(-z_3)\theta(-\gamma) \mathscr{I}^\mathrm{a}_\nu(\bp_\perp;0,-\sqrt{-\gamma}) \,.
    \label{eqapp:jadecomp}
\end{align}
In addition, from the definitions in Eqs.~\eqref{eq:Jmpdef} and \eqref{eq:Jadef}, the symmetry relations are manifest as
\begin{subequations}
\begin{align}
    \mathscr{I}^{(\mp)}_\nu(\bp_\perp;z_0,z_3) &= \mathscr{I}^{(\pm)}_\nu(\bp_\perp;-z_0,z_3) \,, \\
    \mathscr{I}^{(\mp)}_\nu(\bp_\perp;z_0,z_3) &= \mathscr{I}^{(\mp)}_\nu(\bp_\perp;z_0,-z_3) \,, \\
    \mathscr{I}^\mathrm{a}_\nu(\bp_\perp;z_0,z_3) &= \mathscr{I}^\mathrm{a}_\nu(\bp_\perp;-z_0,z_3) \,,
\end{align}
\end{subequations}
and they are useful to simplify the calculations.  In particular, $\mathscr{I}^{(+)}_\nu$ can be retrieved from $\mathscr{I}^{(-)}_\nu$, and thus, we discuss only $\mathscr{I}^{(-)}_\nu$ and $\mathscr{I}^\mathrm{a}_\nu$ below.

We now transform the integrals that appear in Eqs.~\eqref{eqapp:jmpdecomp} and \eqref{eqapp:jadecomp} by using the Weber parabolic cylinder functions~\cite{Gradshteyn:1943cpj} represented by
\begin{align}
    D_\nu[(1+\rmi)\theta] = \frac{2^{-1-\frac{\nu}{2}}}{\sqrt{\pi}} \, \exp\Bigl[ -\rmi\frac{\pi}{4}(\nu+1) + \rmi\frac{\theta^2}{2} \Bigr] \int_{-\infty}^{+\infty} (\rmi x)^\nu \, \rme^{\rmi\frac{x^2}{4} - \rmi x\theta}\, \rmd x \,,
    \label{eqapp:Weber-integ-formula1}
\end{align}
which is immediately derived from the integral representation~\eqref{eq:D-def} for $\theta\in\mathbb{R}$ and $\Re(\nu)>-1$.  We note that $(\rmi x)^\nu$ has a branch cut and in our convention $\mathrm{arg}(\rmi x)^\nu =(\pi/2)\nu\,\mathrm{sgn}(x)$ is chosen in the above integrand.  Also, for $\Re(\nu) < 0$, we can use
\begin{align}
    \int_{-\infty}^{+\infty} (\rmi x)^\nu \,  \rme^{\rmi xy}\, \rmd x = \frac{2\pi}{\Gamma(-\nu)}y^{-\nu-1} \, \theta(y) \,.
    \label{eqapp:Weber-integ-formula2}
\end{align}
After tedious but straightforward calculations, we finally reach
\begin{align}
    \mathscr{I}^{(-)}_\nu(\bp_\perp;-\sqrt{\gamma},0) = \frac{eE\sqrt{\pi}}{\Gamma(-\nu)} \, \rme^{-\rmi\frac{\pi}{4}} \, \int\limits_{\Gamma_1+\Gamma_2} \rmd s \, I_\nu(s,\gamma)
\end{align}
for $z_0 < 0$ and $\gamma>0$, and otherwise,
\begin{align}
   \mathscr{I}^{(-)}_\nu(\bp_\perp;z_0, z_3) = \frac{eE\sqrt{\pi}}{\Gamma(-\nu)} \, \rme^{-\rmi\frac{\pi}{4}} \int\limits_{\Gamma_\mathrm{c}} \rmd s \, I_\nu(s,\gamma) \,.
\end{align}
For the definition of $I_\nu(s)$, Eq.~\eqref{eq:I-def} should be referred to.
In the same way,
\begin{align}
    \mathscr{I}^\mathrm{a}_\nu(\bp_\perp;0,\sqrt{-\gamma}) = \frac{eE\sqrt{\pi}}{\Gamma(-\nu)} \, \rme^{-\rmi\pi\nu -\rmi\frac{\pi}{4}} \int\limits_{\Gamma_1+\Gamma_4} \rmd s \, I_\nu(s,\gamma)
\end{align}
for $z_3>0$ and $\gamma<0$.  For other regions, we have
\begin{align}
    \mathscr{I}^\mathrm{a}_\nu(\bp_\perp;z_0,z_3) = \frac{eE\sqrt{\pi}}{\Gamma(-\nu)} \, \rme^{-\rmi\pi\nu-\rmi\frac{\pi}{4}} \int\limits_{\Gamma_\text{c}} \rmd s \, I_\nu(s,\gamma) \,.
\end{align}
These expressions look similar except for the integration paths with respect to $s$.  We note, however, that the phase factor, $\rme^{-\rmi \pi \nu}$, is crucially important because it contributes to the transverse momentum integration; see Eqs.~\eqref{eq:D0>J}, \eqref{eq:D0<J}, and \eqref{eq:DaJ} and $\nu_\text{s}$ below them.

Because of the applicability of Eqs.~\eqref{eqapp:Weber-integ-formula1} and \eqref{eqapp:Weber-integ-formula2}, these expressions are \textit{a priori} valid only for $-1<\Re(\nu)<0$.  We should be careful when we apply these results to the Dirac field expressions that involves both edges at $\Re(\nu)=-1$ and $\Re(\nu)=0$ (see Sec.~\ref{sec:outline}).  Fortunately, however, it is known that these integral representations remain valid for $-1\leq\Re(\nu)\leq0$ \textit{a posteriori}~\cite{Schwinger:1951nm,Nikishov:1969tt}. Substituting these into Eqs.~\eqref{eqapp:jmpdecomp} and \eqref{eqapp:jadecomp} yields the expressions in Eqs.~\eqref{eq:Jmp_prop} and \eqref{eq:Ja_prop}.

In the derivation above, we implicitly assume that either $z_0$ or $z_3$ is nonvanishing. 
Actually, the point $z_0=0$ and $z_3=0$ is singular, and we need to check whether the integral representations derived above are applicable at this singular point. 
According to Ref.~\cite{Fradkin:1991zq}, the integral representations hold valid after careful treatments.  The justification is straightforward but technically complicated with some identities with respect to the delta functions.

%%%%%%%%%%
\subsection{Propagators in sQED}
\label{app:proptime-sQED}

We now show how to rewrite the propagators in proper-time representation, treating the scalar and Dirac fields in turn. In what follows, to ensure convergence of the integrals at infinity on the $s$-plane, we add an $i\epsilon$ term as $m^2\to m^2-\rmi\epsilon$.
We summarize the integration contours in Fig.~\ref{fig:path}, 
treating the integrals at $s=0$ and $s=-\rmi\pi/(eE)$ as principal values. 

Let us start to derive the proper-time representation of $D^{(-)}_0(x,y)$. 
Substituting \eqref{eq:Jmp_prop} into \eqref{eq:D0>J}, we obtain 
\begin{align}
    D^{>}_0(x,y)
    &= \frac{eE}{16\pi^3} \, \rme^{\frac{\rmi eE}{2}z_3(x_0+y_0)} \biggl\{ \bigl[\theta(z_0) + \theta(-z_0)\theta(-\gamma) \bigr] \int\limits_{\Gamma_\mathrm{c}} \! \rmd s \, 
    + \theta(-z_0)\theta(\gamma) \int\limits_{\Gamma_1+\Gamma_2} \! \rmd s\,  \biggr\} \times\notag\\
    &\qquad \times \frac{\rme^{eEs}}{\sinh{(eEs)}} \, \rme^{-\frac{\rmi eE}{4}(z_0^2-z_3^2)\coth(eEs)} \intperp \, \rme^{\rmi\bp_\perp\cdot \bm{z}_\perp}e^{(2eE\nus)s }. \label{app:D^-integ}
\end{align}
The remaining momentum integral is elementary and can be performed as 
\begin{align}
    \intperp \, \rme^{\rmi\bp_\perp \cdot \bz_\perp} \, \rme^{(2eE\nus) s} = \frac{\pi}{s} \, \rme^{-\rmi\frac{\pi}2} \, \rme^{\rmi\frac{\bz_\perp^2}{4s}} \, \rme^{-eEs} \, \rme^{-\rmi(m^2-\rmi\epsilon)} \,.
    \label{app:D^-gaussint}
\end{align}
Here, we use the definition of $\nus$ given in Eq.~\eqref{eq:nus}. 
From Eqs.~\eqref{app:D^-integ} and \eqref{app:D^-gaussint}, we finally obtain 
\begin{align}
    D^{>}_0(x,y)
    =-\rmi\biggl\{ \theta(-z_0)\theta(z_0^2\!-\!z_3^2) \int\limits_{\Gamma_1+\Gamma_2} \!\!\! \rmd s \, f(s;x,y) + \bigl[ \theta(z_0)+\theta(-z_0)\theta(z_3^2\!-\!z_0^2) \bigr] \int\limits_{\Gamma_\mathrm{c}} \rmd s \, f(s;x,y) \biggr\} \,,
\end{align}
where $f(s;x,y)$ is defined in Eq.~\eqref{eq:f}. 
Performing the similar calculation, we can easily obtain the proper-time representation for $D^{<}_0(x,y)$ as 
\begin{align}
    D^{<}_0(x,y)
    = -\rmi\biggl\{ \theta(z_0)\theta(z_0^2\!-\!z_3^2) \int\limits_{\Gamma_1+\Gamma_2} \!\!\! \rmd s \, f(s;x,y) + [\theta(-z_0)+\theta(z_0)\theta(z_3^2\!-\!z_0^2)] \int\limits_{\Gamma_\mathrm{c}} \rmd s \, f(s;x,y) \biggr\} \,.
\end{align}
The derivation of $D^\mathrm{a}(x,y)$ proceeds in essentially the same way with some care, and we briefly explain details here.  
From Eqs.~\eqref{eq:DaJ} and \eqref{eq:Ja_prop}, we easily obtain
\begin{align}
    \Da(x,y)
    &= -\frac{eE}{16\pi^3} \, \rme^{\frac{\rmi eE}{2}z_3(x_0+y_0)} \biggl\{ \bigl[ \theta(-z_3) + \theta(z_3)\theta(\gamma) \bigr] \int\limits_{\Gamma_\mathrm{c}} \! \rmd s \, 
    + \theta(z_3)\theta(-\gamma) \int\limits_{\Gamma_1+\Gamma_4} \rmd s\, \biggr\} \times\notag\\
    &\qquad\times \frac{\rme^{eEs}}{\sinh{(eEs)}} \, \rme^{-\rmi\frac{eE}{4}(z_0^2-z_3^2)\coth(eEs)} \intperp \, \rme^{\rmi\bp_\perp\cdot \bz_\perp} \, \rme^{2eE\nus(s-\frac{\rmi\pi}{eE})} \,. \label{app:Dainteg}
\end{align}
For the convenience, we change the integration variable as $s\to s-\frac{\rmi\pi}{eE}$, and obtain
\begin{align}
    \Da(x,y)
    = \rmi \biggl\{ \theta(z_3)\theta(z_3^2-z_0^2) \!\! \int\limits_{\Gamma_2+\Gamma_3} \!\!\! \rmd s \, f(s;x,y) + \bigl[ \theta(-z_3) + \theta(z_3)\theta(z_0^2-z_3^2) \bigr] \int\limits_{\Gamma_\mathrm{a}} \! \rmd s \, f(s;x,y) \biggr\} \,.
\end{align}
It should be noted that the integration contours are modified from those of Eq.~\eqref{app:Dainteg} due to the change of the integration variable.

For latter convenience, we rewrite these equations by using the following identities that hold when the integrals are treated as the principal-value integral and when the integration contours are shrunk to paths near the singular points as explicitly indicated in Fig.~\ref{fig:path}, i.e.,
\begin{align}
    & \theta(z_3^2-z_0^2) \int\limits_{\Gamma_\mathrm{c}-\Gamma_1-\Gamma_2} \rmd s \, f(s;x,y) = 0 \,,
    \label{eq:fds=0at0} \\
    & \theta(z_0^2-z_3^2) \int\limits_{\Gamma_\mathrm{a}-\Gamma_2-\Gamma_3} \rmd s \, f(s;x,y) = 0 \,.
    \label{eq:fds=0atipi}
\end{align}
The integration contours are shown in Fig.~\ref{fig:path}.  We can readily see that the contributions from the semicircles are exponentially suppressed depending on the sign of $z_3^2-z_0^2$, and the contribution from a part of the contour parallel to the real axis also vanishes in the infinitesimal limit of the contour. Using Eqs.~\eqref{eq:fds=0at0} and \eqref{eq:fds=0atipi}, we obtain
\begin{align}
    & D^{>}_0(x,y) = -\rmi \biggl\{ \int\limits_{\Gamma_\mathrm{c}} \rmd s \, f(s;x,y) - \theta(-z_0) \int\limits_{\Gamma_\mathrm{c}-\Gamma_1-\Gamma_2} \rmd s \, f(s;x,y) \biggr\} \,,\\
    & D^{<}_0(x,y) = -\rmi \biggl\{ \int\limits_{\Gamma_\mathrm{c}} \rmd s \, f(s;x,y) - \theta(z_0) \int\limits_{\Gamma_\mathrm{c}-\Gamma_1-\Gamma_2} \rmd s \, f(s;x,y) \biggr\} \,,\\
    & \Da(x,y) = \rmi \biggl\{ \theta(z_3)\int\limits_{\Gamma_2+\Gamma_3} \rmd s \, f(s;x,y) + \theta(-z_3) \int\limits_{\Gamma_\mathrm{a}} \rmd s \, f(s;x,y) \biggr\} \,.
\end{align}
Performing the simple algebra, we can easily recover Eq.~\eqref{eq:D++--final}. 

%%%%%%%%%%
\subsection{Propagators in QED}
\label{app:proptime-QED}

As in the scalar case, after substituting Eq.~\eqref{eq:Jmp_prop} into Eq.~\eqref{eq:S0>-J} and Eq.~\eqref{eq:Ja_prop} into Eq.~\eqref{eq:SaJ}, and then performing the integral over $\bp_\perp$, we obtain
\begin{align}
    S^{>}_0(x,y)
    &= \rmi(\rmi\slashed{D}_x+m) \biggl\{ \theta(-z_0)\theta(z_0^2-z_3^2) \int\limits_{\Gamma_1+\Gamma_2} \rmd s\, g(s;x,y) + \notag\\
    &\qquad\qquad\qquad + \bigl[ \theta(z_0)+\theta(-z_0)\theta(z_3^2-z_0^2) \bigr] \int\limits_{\Gamma_\mathrm{c}} \rmd s \, g(s;x,y) \biggr\} \,,\\
    S^{<}_0(x,y)
    &= -\rmi(\rmi\slashed{D}_x+m) \biggl\{ \theta(z_0)\theta(z_0^2-z_3^2) \int\limits_{\Gamma_1+\Gamma_2} \rmd s \, g(s;x,y) + \notag\\
    &\qquad\qquad\qquad + \bigl[ \theta(-z_0)+\theta( z_0)\theta(z_3^2-z_0^2) \bigr] \int\limits_{\Gamma_\mathrm{c}} \rmd s \, g(s;x,y) \biggr\} \,,\\
    \Sa(x,y)
    &= -\rmi(\rmi\slashed{D}_x+m) \biggl\{ \theta(z_3)\theta(z_3^2-z_0^2) \int\limits_{\Gamma_2+\Gamma_3} \rmd s \, g(s;x,y) + \notag\\
    &\qquad\qquad\qquad + \bigl[ \theta(-z_3)+\theta(z_3)\theta(z_0^2-z_3^2) \bigr] \int\limits_{\Gamma_\mathrm{a}} \rmd s \, g(s;x,y) \biggr\} \,,
\end{align}
where the definition of $g(s;x,y)$ is given above Eq.~\eqref{eq:f}.
The relative sign difference between $S^>(x,y)$ and $S^<(x,y)$ is a consequence of the anticommutative nature of fermions.  
By using the following equations that are valid under the same conditions for Eqs.~\eqref{eq:fds=0at0} and \eqref{eq:fds=0atipi},
\begin{align}
    &\theta(z_3^2-z_0^2)  \int\limits_{\Gamma_\mathrm{c}-\Gamma_1-\Gamma_2} \rmd s \, g(s;x,y) = 0 \,,
    \label{eq:gds=0at0} \\
    &\theta(z_0^2-z_3^2) \int\limits_{\Gamma_\mathrm{a}-\Gamma_2-\Gamma_3} \rmd s \, g(s;x,y) = 0 \,.
    \label{eq:gds=0atipi}
\end{align}
The integration contours are shown in Fig.~\ref{fig:path}.
Then, we can rewrite the propagators as
\begin{align}
    & S^{>}_0(x,y)
    = \rmi(\rmi\slashed{D}_x+m) \Biggl[ \;\; \int\limits_{\Gamma_\mathrm{c}} \rmd s \, g(s;x,y) - \theta(-z_0) \int\limits_{\Gamma_\mathrm{c}-\Gamma_1-\Gamma_2} \!\! \rmd s \, g(s;x,y) \Biggr] \,,\\
    & S^{<}_0(x,y)
    = -\rmi(\rmi\slashed{D}_x+m) \Biggl[ \;\; \int\limits_{\Gamma_\mathrm{c}} \rmd s \, g(s;x,y) - \theta(z_0) \int\limits_{\Gamma_\mathrm{c}-\Gamma_1-\Gamma_2} \!\! \rmd s \, g(s;x,y) \Biggr] \,,\\
    & \Sa(x,y)
    = -\rmi(\rmi\slashed{D}_x+m) \Biggl[ \theta(z_3)\int\limits_{\Gamma_2+\Gamma_3} \! \rmd s \, g(s;x,y) + \theta(-z_3)\int\limits_{\Gamma_\mathrm{a}} \rmd s \, g(s;x,y) \Biggr] \,.
\end{align}
From these representations, we can easily recover Eq.~\eqref{eq:S++--final}.

\bibliographystyle{JHEP}
\bibliography{main}

\end{document}